\documentclass[sn-apa]{sn-jnl}

\usepackage{url}
\usepackage{CJKutf8}
\usepackage{multirow}
\usepackage{colortbl}
\usepackage{subcaption}
\usepackage{marvosym}
\usepackage{etoolbox}
\usepackage[dont-mess-around]{fnpct}
\usepackage{graphicx}

\jyear{2022}%
\begin{document}

\title[Exploring the Confounding Factors of Academic Career Success]{Exploring the Confounding Factors of Academic Career Success: An Empirical Study with Deep Predictive Modeling}

\author[1]{\fnm{Chenguang} \sur{Du}}\email{duchenguang@buaa.edu.cn}
\author*[1]{\fnm{Deqing} \sur{Wang}}\email{dqwang@buaa.edu.cn}
\author[2]{\fnm{Fuzhen} \sur{Zhuang}}\email{zhuangfuzhen@buaa.edu.cn}
\author[3]{\fnm{Hengshu} \sur{Zhu}}\email{zhuhengshu@gmail.com}

\affil*[1]{\orgdiv{School of Computer}, \orgname{Beihang University}, \orgaddress{\street{Xueyuan Road}, \city{Haidian District}, \postcode{100191}, \state{Beijing}, \country{China}}}
\affil[2]{\orgdiv{Institute of Artificial Intelligence}, \orgname{Beihang University}, \orgaddress{\street{Xueyuan Road}, \city{Haidian District}, \postcode{100191}, \state{Beijing}, \country{China}}}
\affil[3]{\orgdiv{Baidu Talent Intelligence Center}, \orgname{Baidu Inc.}, \orgaddress{\street{Shangdi 10th Street}, \city{Haidian District}, \postcode{100085}, \state{Beijing}, \country{China}}}

\abstract{Understanding determinants of success in academic careers is critically important to both scholars and their employing organizations. While considerable research efforts have been made in this direction, there is still a lack of a quantitative approach to modeling the academic careers of scholars due to the massive confounding factors. To this end, in this paper, we propose to explore the determinants of academic career success through an empirical and predictive modeling perspective, with a focus on two typical academic honors, i.e., IEEE Fellow and ACM Fellow. We analyze the importance of different factors quantitatively, and obtain some insightful findings. Specifically, we analyze the co-author network and find that potential scholars work closely with influential scholars early on and more closely as they grow. Then we compare the academic performance of male and female Fellows. After comparison, we find that to be elected, females need to put in more effort than males. In addition, we also find that being a Fellow could not bring the improvements of citations and productivity growth. We hope these derived factors and findings can help scholars to improve their competitiveness and develop well in their academic careers.}


\maketitle

\section{Introduction}
\label{intro}
Academic career success is the pursuit of every scholars. Exploring determinants of success in academic careers can not only help scholars to develop themselves better, but also guide their employing organizations to evaluate and manage talents scientifically.
Recently, considerable researchers have explored the impact of various factors on academic careers, such as co-author network~\citep{wu2020meta,li2019early} and scientific impact~\citep{nie2019academic,Way2017}, etc.
However, scholars' academic careers are affected by massive confounding factors. It is challenging to modeling the academic careers of scholars with a quantitative and systematic approach.
To this end, in this paper, we try to explore the determinants of academic career success through an empirical and predictive modeling perspective. In particular, we focus our study on IEEE Fellow and ACM Fellow which are two typical career honors.

\begin{figure}
\centering
\includegraphics[width=0.8\columnwidth]{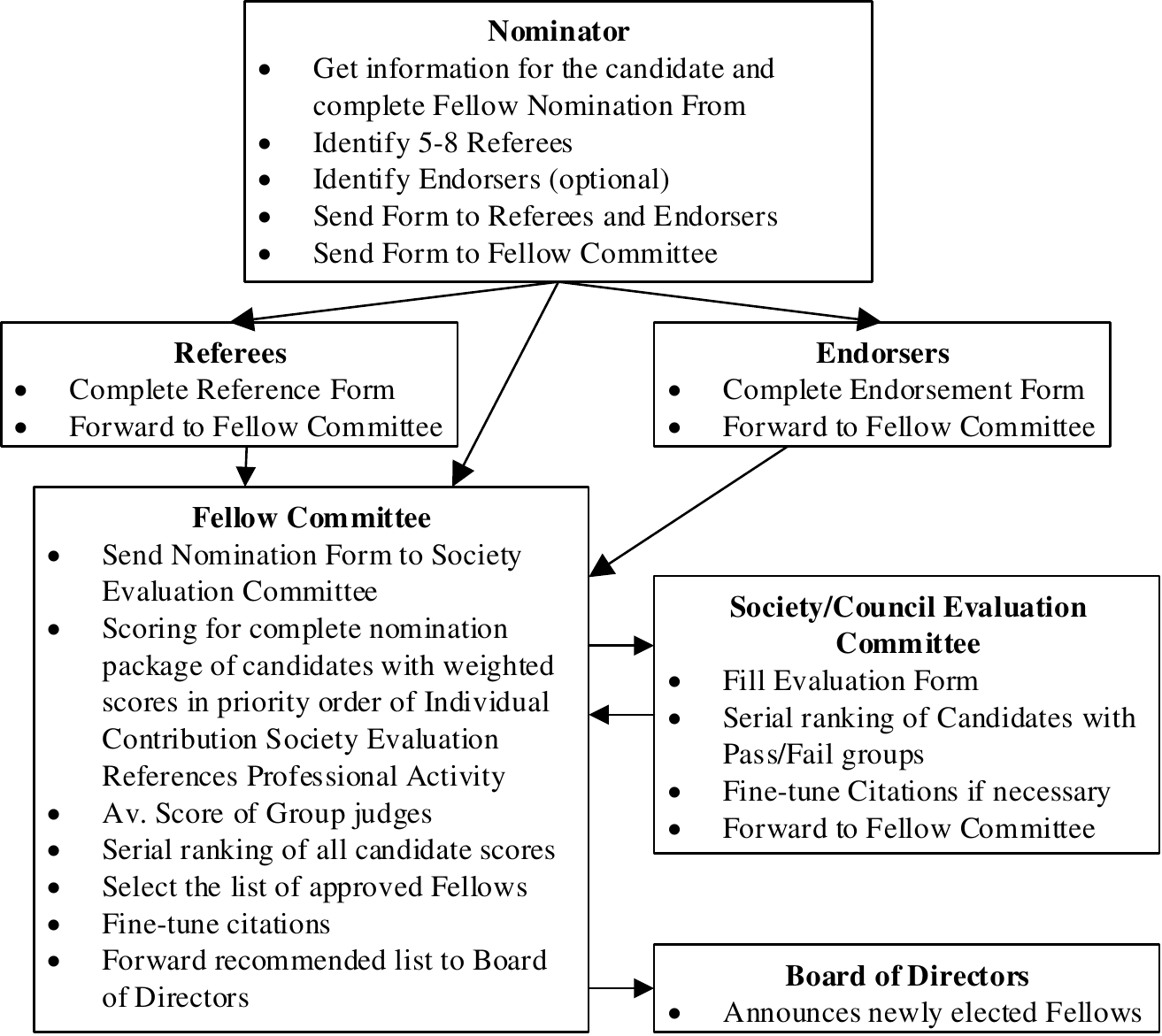}
\caption{Flowchart for IEEE Fellowship election~\citep{bimal05}.}
\label{fig:flowchart}
\end{figure}

Taking the IEEE Fellow as an example, to be elected as a Fellow, the candidate must have accrued a sustained level of contribution over time with clear impact that extends well beyond his/her own organization, and must secure endorsements from 5-8 IEEE members, preferably individuals who are themselves IEEE Fellows or have otherwise achieved distinction in the field ~\citep{bimal05}. Specifically, the candidate will submit a nomination, which includes 1) his/her educational background, 2) his/her most significant professional accomplishments and their foundational, technical, commercial, or other achievements, and 3) his/her most significant leadership roles and awards in IEEE or other service activities. Moreover, the candidate must recommend 5-8 Referees to write supporting letters for her. Then the Fellow Committee will rate each nominee numerically on the basis of the above information and recommend nominees according to some criteria. The detailed flowchart for IEEE Fellowship election ~\citep{IEEE18} is shown in Fig.~\ref{fig:flowchart}.

However, after submitting the nominations, the candidates do not know whether and how they are elected as a Fellow, because the  evaluation criteria and rating process are complicated, and even secret to candidates. Moreover, the evaluation of Fellow Committee is a combined process between subjective evaluation and objective evaluation, because the members of Fellow Committee (or Society Evaluation Committee) often determine their recommendation score by their professional knowledge and some objective metrics, e.g., scientific impact, contribution to the community, citations, etc. 
To this end, we try to search for the answers of the following challenging questions from a data-driven perspective.

 \textbf{Question I:  What factors are crucial for a candidate to be elected as an ACM/IEEE Fellow and how can we use these factors to represent the candidate quantitatively and objectively?}

 \textbf{Question II: How to mine the key factors of candidates by modeling the Fellow election? } 

That is, we need to build a classification model to predict whether a candidate will be elected as Fellow at current year according to his/her scholarly outputs. If not, a regression model should be built to predict how many additional years it will take a candidate to be a Fellow.

For the first problem, we observe that a candidate will provide his/her personal educational background (e.g., institution, degree, year conferred, and major discipline) and the professional accomplishments by submitting his/her nomination~\citep{Petersen2014}. Even though the content in his/her nomination are not available to us, we can collect the related data from his/her public venues, such as personal website, Google Scholar, etc. For professional accomplishments, many studies~\citep{Petersen2014,Sinatra2016,Way2016} often quantitatively evaluate a researcher's professional accomplishments from his/her scholarly productivity ~\citep{Way2016,Sinatra2016,Way2017} and scientific impact~\citep{Penner2013}. The above factors are not only valuable for modeling the professional accomplishments of a candidate but also very important to Fellow election, because these factors are practically evaluated by Fellow Committee members during the rating process of nominations. However, in addition to the above, there is still one challenge for predicting whether a candidate can be elected as an ACM/IEEE Fellow in a specific year. Specifically, the challenge is that the election of ACM/IEEE Fellow involves to recommend the endorsers from existing Fellows to write supporting letters, which means whether the candidate can receive good evaluations from authoritative peer-reviews. In practical terms, it is difficult for us to obtain the endorsers' names of each candidate.  In this paper, we propose to measure this criteria quantitatively from co-author networks with the help of node2vec~\citep{Grover2016}. That is, we propose and define a ``scholarly distance" between a candidate and the existing Fellows to measure the ability that a candidate can recommend appropriate endorsers to write endorsements for herself. Details about scholarly distance can be found in Section~\ref{data}. 

For the second problem, we have the annual scholarly outcomes of each Fellow from the year of his/her first publication and we can represent each Fellow as time sequence vectors by collecting his/her scholarly outcomes. It is worthy noting that the election criteria of ACM/IEEE Fellow is changing over time because the candidates have to face severe global competition. For instance, the evaluation criteria of Fellow candidates in 1980 is totally different with that in 2016. Therefore, our model should be time-related and dynamic, and can be explained from two aspects. First, the nomination becomes more fiercer with the rapid growth of talented candidates, thus the criteria becomes higher and higher. It means that our model should pay more attention to the latest data of recent elected Fellows, because the latest data indicates more precise criteria of Fellow election. Second, a candidate will have higher probability of being elected as a Fellow if he/she spends more years on research, that is, the accumulation of his/her scholarly outcomes and impacts will help his/her to be more competitive. To address the above problems, we design a base neural network structure with self-attention mechanism. By connecting different output layers, we construct a classification model (named Cls-Fellow model) and a regression model (named Reg-Fellow model) to answer Question (II). Extensive experiments show that our proposed models can achieve great performance.

This is the first work that focuses on data-driven method to mine the key factors for Fellow career path. Our main research contributions are summarized as follows.

\begin{itemize}
  \item We analyze the factors of being an ACM/IEEE Fellow quantitatively and propose the definition of ``scholarly distance'' to measure the ability that a candidate can recommend appropriate endorsers for herself.
  \item We propose two self-attention based neural network  models to predict whether a Fellow candidate will be elected as a Fellow at current year and how many additional years it will take his/her if he/she could not be named Fellow, which can be used for scholars' self-assessment.
  \item Based on our datasets and models, we discuss the importance of different factors in academic career. Moreover, we discover some interesting phenomena, e.g., the evolution of co-author networks between candidates and Fellows, the inequality of gender, the reputation changes before and after being a Fellow, etc. These findings could be helpful for scholars to improve their competitiveness and develop well in their academic career.
\end{itemize}

\section{Related Work}
In this section, we review two categories of literature that are related to this paper, namely research on quantitative analysis of career trajectory of scientists, and research on data mining for talents.

\subsection{Quantitative analysis of career trajectory of scientists.} 
How to evaluate a scientist's professional accomplishments quantitatively is a key question during his/her career trajectory. Recently, many studies try to handle it by considering scientists' scholarly productivity ~\citep{Petersen2014,Sinatra2016,Way2016} and scientific impact ~\citep{Penner2013,Way2016,Sinatra2016,Way2017} from data perspective. For example, ~\citet{Petersen2012} modeled the productivity and fluctuations over the academic career and found the persistence and uncertainty in the academic community. ~\citet{Way2017} demonstrated that  2/3 faculties exhibited a rich diversity of productivity  patterns, rather than simple ``rise-down'' pattern. ~\citet{nie2019academic} identified the academic rising star by using the increment of a scholar’s comprehensive evaluation score and a non-iterative hierarchical citation-based model. ~\citet{min2021identifying} introduced a perspective of dynamic citation process to identify citation patterns of scientific breakthroughs. Besides, some papers study the inequality of gender, knowledge in program review, peer review, hiring network, etc. For instance, ~\citet{Kevin2016} proposed the intelligent distance affected program review, i.e., reviewers in the same research field tend to give a lower score for program review. ~\citet{Ginther2011} reported the inequality of race and ethnicity in NIH research awards. \citet{Clausete2015,Way2016,Way2017} showed the inequality of gender in peer review and hiring network of the faculty.

\subsection{Data mining for talents.} 
Recently, how to use data mining techniques to address human resource management have attracted researchers' much attention in data mining and machine learning communities~\cite{Zhu2016,Li2017}. For example, ~\citet{Xu2014} showed a novel method for modeling the professional similarity by mining professional career trajectories. ~\citet{de2018human} proposed an analytical method to generate an intelligible classification of job roles and skill sets.  ~\citet{li2017prospecting} proposed a survival analysis approach to model the talent career paths with a focus on turnover and career progression. Based on graph neural network, ~\citet{ye2019identifying} proposed to utilize Graph Convolutional Network~\cite{defferrard2016convolutional} to extract the local information of employees in their organizational social network for high-potential talent identification and ~\citet{wu2020meta} propose a  Mate-path Hierarchical Heterogeneous Graph Convolution Network for high-potential scholar recognition.

Differing from the above studies, we not only try to analyze the factors of Fellow career trajectory quantitatively but also propose both classification and regression models to predict whether a candidate will be elected as a Fellow at current year and how many additional years it will take he/she to be a Fellow.

\section{Data and factors}
\label{data}
Up to the date (10/26/2020), there are 10,483 IEEE Fellows (including 425 females) and 1,221 ACM Fellows (including 155 females), spanning from 1934 to 2020 for IEEE Fellow and from 1994 to 2019 for ACM Fellow.  First, we collected the basic information (e.g., name, region, country, gender, etc) of each Fellow from IEEE website\footnote{https://www.ieee.org/membership/fellows/fellows-directory.html} and ACM website\footnote{https://awards.acm.org/award\_winners}. Second, we collected their publications (including title, authors, journal, year, etc) from Microsoft Academic Search\footnote{https://academic.microsoft.com} and  the corresponding citation metrics (such as citations of each paper, $h$-index, $i10$-index) from Google Scholar\footnote{https://scholar.google.com}. Then we removed the Fellows who had no more than 300 citations or were elected less than 8 years since his/her first publication because of noisy data. Fourth, we also collected scholars in ACM Distinguished Members and Aminer Highly cited Scholars Library who are not IEEE/ACM Fellows as negative examples (named non-Fellow) for our classification model. We collected non-Fellow data (same field as Fellows) from ACM website\footnote{https://awards.acm.org/distinguished-members/award-winners} and Aminer website\footnote{https://gct.aminer.cn/eb/series?name=\begin{CJK}{UTF8}{gbsn}高引学者系列\end{CJK}[2016]}. Finally, we obtained three datasets consisting of 7,191 (877 female) talented researchers with 1,377,907 publications.
Some basic statistics  of the three datasets are shown in Table~\ref{tab:data}.

\begin{table}
\newcommand{\tabincell}[2]{\begin{tabular}{@{}#1@{}}#2\end{tabular}}
\caption{The details of the Fellow datasets}
\label{tab:data}
\begin{minipage}{\columnwidth}
\begin{center}
\begin{tabular}{cc|cccc}
\hline
\textbf{\tabincell{c}{Fellow.\\ Dataset}} &\textbf{Source} & \textbf{Male} & \textbf{Female} &\textbf{ \tabincell{c}{Avg.\\ Pubs}} & \textbf{\tabincell{c}{Avg. \\cites}} \\
\hline
ACM & ACM Fellow   & 820 & 121  & 143.5  & 10,388.8\\
IEEE & IEEE Fellow   & 4,044 & 275  & 145.7   & 4,322.8\\
non  & ACM DIST.  & 363 & 54  & 151.0   & 6,952.0\\
non  & Aminer & 1,135 & 427  & 139.1   & 10,397.9\\
\hline
\end{tabular}
\end{center}
\end{minipage}
\end{table}

After checking the material of Fellow nomination carefully and considering the factors mentioned in Reference~\citep{bimal05}, we describe the important factors used in our model and provide some statistical information as follows to answer the Question (I).

\textbf{Accumulation time.} How many years does it cost a talented scientist to achieve his/her academic accumulation and to be elected as an ACM/IEEE Fellow after his/her first publication? In this paper, we define the years as $t_i=ey_i-sy_i$, where $ey_i$ is the year when Fellow $i$ in our datasets was elected as an ACM/IEEE Fellow and $sy_i$ is the year when his/her first paper was published.

\textbf{Scholarly productivity.} Similar to previous studies~\citep{Petersen2012,Sinatra2016}, we model the academic career trajectory as a sequence of scientific outcomes which arrive at the variable rate $n_i(t)$. Here $n_i(t)$ is the annual productivity of Fellow $i$ at $t-$th year after his/her first publication. Generally, the reputation of a scientist is typically a cumulative representation of his/her contributions, we consider the cumulative production $N_i(t)=\sum_{t'=1}^{t}n_i(t')$ as a proxy for career achievement. $N_i(t)$ is the total number of papers Fellow $i$ publishes up to time $t$ after his/her first publication, which asymptotically follows $N_i(t)\sim t^{\alpha_i}$. The $\alpha_{i}$ quantifies the career trajectory dynamics and $\alpha_{i}> 1$ indicates on average a steady increase in his/her productivity with $t$. 
Note that productivity is only one metric in our model and higher productivity is not equal to higher impact. For scholarly productivity of Fellow $i$, we consider 4 factors:  annual publications $n_i(t)$, annual average publications $\bar{n}_i(t)$, the total number of publications $N_i(cy)$ and $\alpha_{i}(t)$.

\textbf{Scientific impact.} Existing measures of scientific impact believe that citations offer a quantitative proxy of the importance of findings or a scientist's standing in the research community~\citep{BORNMANN2011,Penner2013}. Like previous studies, we take the total number of citations and annual citations~\citep{Penner2013} into account. Besides, $h$-index~\citep{BORNMANN2011} and $i10$-index are also widely used metrics for the evaluation of scientific impact. The $h$-index incorporates productivity as well as citation impact, and $i10$-index is the number of publications with at least 10 citations~\citep{Google}. Actually, these measures of scientific impact are often debated. In our models, we include above measures and let the model consider the scientific impact of candidates, comprehensively.

\textbf{Scholarly distance.} When submitting his/her nomination, a candidate has to recommend 5-8 existing ACM/IEEE Fellows who are familiar with his/her fields as endorsers. How to measure the social ability that the candidate can recommend appropriate endorsers is one key challenge. Generally, candidates recommend endorsers by their co-author network. To address this challenge, we define a scholarly distance score ${sd}_i$ between a candidate $i$ and existing Fellows by applying co-author network and node2vec~\citep{Grover2016}, as shown in Eq.~\ref{equ:sd}.
\begin{equation}
\label{equ:sd}
sd_i = \frac{\sum_{j=1}^{j=N}cos(\mathbf{x}_i,\mathbf{x}_j)}{N},
\end{equation}
\noindent where $\mathbf{x}_i$, $\mathbf{x}_j$ are the low-dimensional graph embedding vectors learned by node2vec~\citep{Grover2016} of candidate $i$ and Fellow $j$ from their co-author networks, respectively; $N$ is the total number of Fellows. Generally, a higher scholarly distance score indicates that the candidate owns a better co-author network with existing Fellows, thus he/she can recommend more appropriate endorsers for his/her nomination.

\textbf{Scholarly circle.} ~\citet{ye2019identifying} proposed a data-driven approach for identifying high potential talents (HIPOs) from the newly-enrolled employees by modeling the dynamics of their behaviors in organizational social networks and they found that HIPOs can promote their social centrality factors more effectively in terms of both speed and numerical value. 

In this paper, we utilize Hierarchical Graph Convolutional Network (GCN)~\citep{defferrard2016convolutional} to extract candidates’ co-author network (graph) information. ~\citet{singh2020evolution} found that author pairs who have a co-authorship distance $d \leq 3$ significantly affect each other's citations, but this effect falls off rapidly for longer distances in the co-author network. Thus, for each candidate, we build co-author graph with his/her co-authors (within co-authorship distance $d \leq 3$) and number of cooperation. In each graph, each node represents a scholar and each edge represents the co-author relationship between two scholars. We label each node with values 0, 1 and 2 representing non-Fellow, Fellow and current candidate, respectively. The weight of each edge is the number of cooperation between two scholars.

\textbf{Field of Research.} Due to the different development trends in various research fields, the difficulty of election is not the same. For example, candidates can be organized by 39 societies which focus on different research fields in IEEE. The larger Societies have more nominations and competition tends to be more severe. Meanwhile, the score tends to be high normally because of less competition in a small Society ~\citep{bimal05}. Owing to the significant impact of the research fields, we define a research field vector ${rf}$ representing candidates' fields. For IEEE Fellow candidates, the field vector ${rf}$ is based on the research fields of 39 IEEE Societies\footnote{https://www.ieee.org/communities/societies/index.html}. And 34 child topics of ``Computer Science'' in Microsoft Academic topics\footnote{https://academic.microsoft.com/topics/41008148} are used to generate the field vector ${rf}$ for ACM Fellow candidates. However, the categories of IEEE Societies and Microsoft Academic topics often overlaps in research fields. Overlapping categories can make it difficult to classify candidates and their publications. Therefore, we combined IEEE Societies and Microsoft topics into 8 broader categories based on the similarity of research fields, independently. For example, the ``Computer graphics'' and the ``Computer vision'' are combined into one broader category. Then, based on conferences and journals related to IEEE Societies and Microsoft Academic topics, we collected 2000 high-cited papers for each of the broad category as a field-related dataset. Finally, we trained a BERT-based ~\citep{devlin2018bert} text classification model (named Field\_Cls) using the field-related dataset we collected. The accuracy of the Field\_Cls model can reach 76.1\% on the ACM categories data and reach 87.0\% on the IEEE categories data. For a candidate $i$, his/her research field vector ${rf}_i$ can be calculated by inputting his/her paper data into the Field\_Cls, as shown in Eq.~\ref{equ:rf}.

\begin{equation}
  \label{equ:rf}
  rf_i = \frac{\sum_{j=1}^{j=N}Field\_Cls(\mathbf{p}_j)}{N},
\end{equation}

\noindent where $N$ is the total number of papers of candidate $i$; $\mathbf{p}_j$ represents the information of a certain paper $j$ which belongs to candidate $i$; $Field\_Cls(\mathbf{p}_j)$, a 8-dimensional vector, is the classification result of the Field\_Cls model on $\mathbf{p}_j$;

\textbf{Place of Employment.} Generally, a reputed employer of the candidate also gives the Fellow Committee members a better first impression~\citep{bimal05}. Therefore, in our models, candidates' employment information is represented using 4-dimensional vector embedding (named $employ\_emb$). 

\textbf{Gender.} The last factor is the gender of candidates. Previous studies show that gender bias~\citep{Leslie2015} exists in academia, such as faculty hiring~\citep{Clausete2015,Way2016}, grant proposal~\citep{vanArensbergen2012} and peer review~\citep{Kaatz2014}. In this paper, our aim is to estimate the influence of gender on the Fellow selection, therefore gender is another factor to be considered in our model.

Finally, for each candidate $i$ in a given calendar year $cy$,  we can represent him/her by a 36-dimensional vector $\mathbf{x}_{i}^{cy}$, with details as follows:
\begin{itemize}
\item \textbf{Accumulation time:} $t_i = cy-sy_i$, where $sy_i$ is the calendar year when candidate $i$ published his/her first paper.
\item \textbf{Gender:} 1 for male and 0 for female.
\item \textbf{Scholarly productivity:}  It has 4 factors. They are the number of publications in the given calendar year ($n_i(cy)$), the $\alpha$ value of publications ($\alpha_i(cy)$), the total number of publications $N_i(cy)$, and the average publications $\overline{n}_i(cy)$ from his/her first publication year to the given calendar year, respectively.
\item \textbf{Scientific impact:} It consists of 5 factors. They are the total citations of Fellow $i$ ($c_i(cy)$),  the $\alpha$ value of citations in the given calendar  year $\alpha ^c_i(cy)$, the average citations $\overline{c}_i(cy)$ from his/her first publication year to the given calendar year, the $h$-index $h_i(cy)$ and $i$10-index $i10_i(cy)$ in the given calendar year, respectively. 
\item \textbf{Scholarly distance:} the current $sd_i(cy)$ in the given calendar year.
\item \textbf{Scholarly circle:} the current co-author network embedding $cn_i(cy)$ with 12-dimensions in the given calendar year.
\item \textbf{Field of Research:} the current $rf_i(cy)$ with 8-dimensions in the given calendar year.
\item \textbf{Place of Employment:} the current $employ\_emb_i(cy)$ with 4-dimensions in the given calendar year.
\end{itemize}
For \textbf{Accumulation time}, \textbf{Gender}, \textbf{Scholarly productivity}, \textbf{Scientific impact} and \textbf{Scholarly distance}, their each dimension is normalized by $z$-score. Finally, the candidate $i$ can be represented as time series vectors $\mathbf{X}_{i}$ from the year of his/her first publication to 2020, as shown in Eq.~\ref{equ:fellow_represent}
\begin{equation}
  \label{equ:fellow_represent}
  \mathbf{X}_{i}=\{\mathbf{x}_{i}^{sy_i},\cdots,\mathbf{x}_{i}^{2020}\},
\end{equation}
\noindent where $sy_i$ is the year of his/her first publication. For example, ``Michael I. Jordan'' started to publish his first paper in 1981, so his vectors are as follows,
$\{\mathbf{x}_{MJ}^{1981},\cdots,\mathbf{x}_{MJ}^{2020}\}$.

In addition, for the above factors, we conduct preliminary statistics and analysis, which can be found in Appendix~\ref{data_analysis}.

\section{Predictive Model}
To answer the Question (II), we need to model the Fellow election: \emph{build a classification model to predict whether a candidate will be elected as a Fellow}, and \emph{build a regression model to predict how many more years it will take a candidate to be a Fellow if he/she could not be named Fellow at current year}.

\begin{figure}[t]
  \centering
  \includegraphics[width=1\columnwidth]{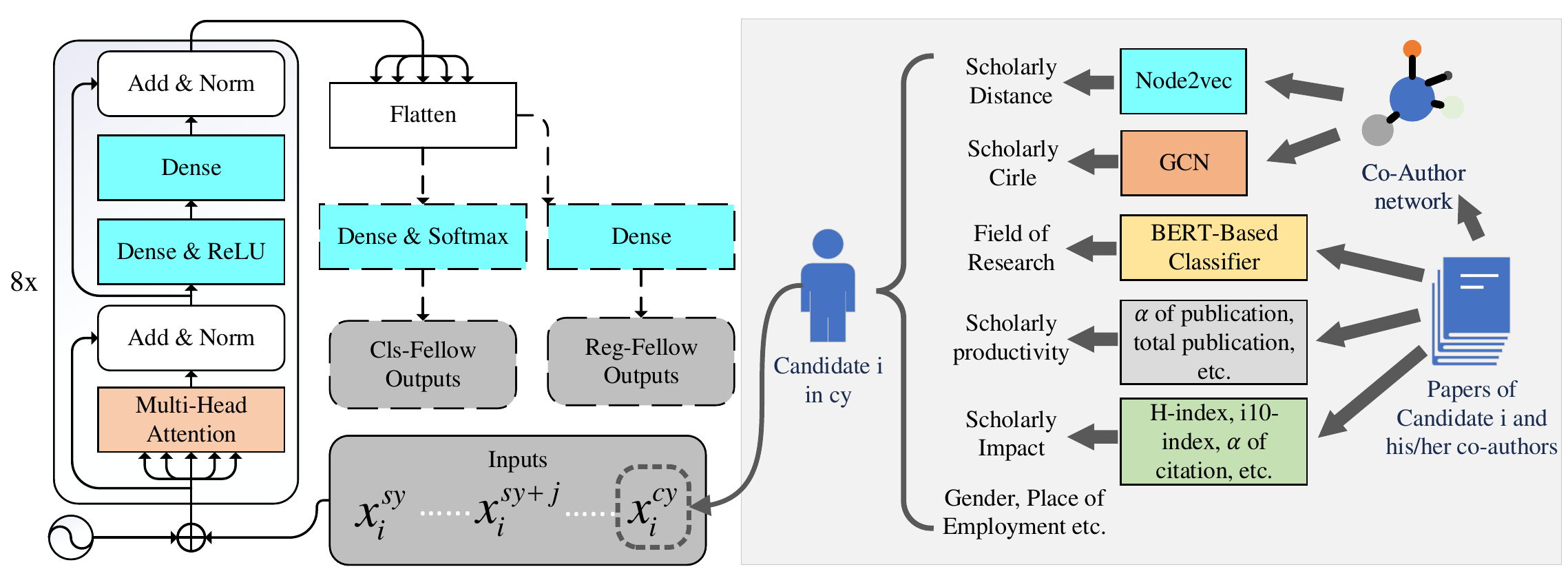}
  \caption{Graphical illustration of the proposed models.}
  \label{fig:model}
\end{figure}

The prediction of the Fellow election is related to and affected by time. Therefore, our models are based on the same underlying neural network structure with multi-head self-attention mechanism. The graphical structure of the proposed models is shown in Fig.~\ref{fig:model}. Noticed the excellent performance of Transformer~\citep{vaswani2017attention} on the seq2seq tasks, multi transformer encode layers are used on the front of the models. By flattening the output of the last encode layer, the information of a candidate can be transformed into one vector as a high-level representation. Finally, different fully connected layers are connected to the flatten layers. We expect these two models are able to capture the correlation between election results and the candidate's performance, as well as the underlying trend and evolving phenomena. 

For co-author network information extraction, two GCN layers are used. By aggregating the output of second GCN layer, the co-author network can be represented by a 12-dimensional vector. After co-author network embedding, we concat co-author network vector and other factor vectors for academic trajectory processing.

\section{Experiments}
Our Cls-Fellow model can predict whether a candidate will be elected as a Fellow in the current year. If candidates are classified as non-Fellow by Cls-model, it means that they may not named Fellow in the current year. For these candidates, our Reg-model can predict how many additional years it will take them to be Fellows.
In this section, we compare our proposed Cls-Fellow model and Reg-model with some state-of-the-art baselines on ACM and IEEE Fellow datasets \footnote{Data and Source Code Repository: \url{https://github.com/nobrowning/Fellow_Analysis}}. During the training and testing, the examples are used in the same way for Cls-Fellow model and the rest baselines. Specifically, the examples are represented as time series vectors, as explained in Section~\ref{data}. The sequence of data is flattened for some baseline methods which cannot take sequence as input.

According to the task type (classification or regression) and Fellow type (ACM or IEEE), we prepare four sub-datasets, namely ACM/IEEE classification datasets and ACM/IEEE regression datasets, respectively. The details of data preparation can be found in Appendix~\ref{exp_prepar}.

\subsection{Baseline Methods}
Multiple transformer encode layers are used in the Cls-Fellow and Reg-Fellow. 
To evaluate the performance of transformer encode layers and the robustness of feature selection, we compared Cls-Fellow and Reg-Fellow with multiple baselines.

For the classification task, 8 kinds of classification models are used as baselines. They are $\epsilon$-Support Vector classifier (SVM)~\citep{Zhou2015ARO}, Linear Regression (LR), Ridge Regression, Decision Tree classifier (DT), Random Forest classifier(RF), Graph Convolutional Network (GCN)~\citep{defferrard2016convolutional}, Multi-layer Perceptron classifier (MLP)~\citep{KingmaB15} and Attention-RNN~\citep{Bahdanau14,xu2018}.

For the regression task, six kinds of regression models are used as baselines. They are Ridge Regression, $\epsilon$-Support Vector Regression (SVR)~\citep{Zhou2015ARO}, Multi-layer Perceptron regressor (MLP)~\citep{KingmaB15}, Decision Tree regressor (DT) and Attention-RNN~\citep{Bahdanau14,xu2018}. Some of them are the same as in classification baselines. 

The implement details of baselines and our models can be found in Appendix~\ref{baseline_detail} and Appendix~\ref{param_setting}, respectively.

\subsection{Results and Analysis}
For classification task, Cls-Fellow model is supposed to classify Fellow or non-Fellow. The F1 scores on IEEE/ACM classification datasets are shown in Table~\ref{tab:acm_cls_result} and Table~\ref{tab:ieee_cls_result}. The results are as follows:

\begin{table} [th]
  \newcommand{\tabincell}[2]{\begin{tabular}{@{}#1@{}}#2\end{tabular}}
  \centering
  \caption{The F1 Score of Classification on ACM Classification Datasets}
  \label{tab:acm_cls_result}
  \resizebox{\columnwidth}{!}{
    \begin{tabular}{c|c|c|c|c|c|c|c|c|c}
    \hline
    &\tabincell{c}{Cls-\\Fellow}&
    \tabincell{c}{Attn-\\RNN}&
    \tabincell{c}{GCN}&
    LR&
    Ridge&
    RF&DT&SVM&MLP\\
    \hline
    2015& 84.5 & 84.1 & 80.3 & 81.7 & 79.9 & 82.6 & 78.8 & 79.2 & 78.1 \\
    2016& 85.2 & 85.3 & 80.9 & 81.6 & 82.7 & 83.9 & 79.1 & 76.9 & 80.7 \\
    2017& 85.3 & 86.1 & 82.2 & 83.2 & 83.4 & 84.4 & 81.8 & 80.3 & 81.3 \\
    2018& 86.6 & 86.7 & 80.9 & 83.8 & 82.7 & 84.9 & 83.4 & 82.1 & 83.0 \\
    2019& 86.4 & 87.2 & 80.8 & 83.7 & 84.6 & 85.4 & 82.8 & 81.1 & 82.1 \\
    \hline
    AVG.& \textbf{85.6} & \textbf{85.9} & 81.0 & 82.8 & 82.7 & 84.2 & 81.2 & 79.9 & 81.0 \\
    \hline
    \end{tabular}
  }
\end{table}

\begin{table} [th]
  \newcommand{\tabincell}[2]{\begin{tabular}{@{}#1@{}}#2\end{tabular}}
  \centering
  \caption{The F1 Score of Classification on IEEE Classification Datasets}
  \label{tab:ieee_cls_result}
  \resizebox{\columnwidth}{!}{
    \begin{tabular}{c|c|c|c|c|c|c|c|c|c}
    \hline
    &\tabincell{c}{Cls-\\Fellow}&
    \tabincell{c}{Attn-\\RNN}&
    \tabincell{c}{GCN}&
    LR&
    Ridge&
    RF&DT&SVM&MLP\\
    \hline
    2015& 82.8 & 82.7 & 79.6 & 77.8 & 81.0 & 81.3 & 79.8 & 75.3 & 75.9 \\
    2016& 83.6 & 83.8 & 78.5 & 77.2 & 80.6 & 81.9 & 81.6 & 74.8 & 74,7 \\
    2017& 83.6 & 84.1 & 78.7 & 75.9 & 81.9 & 82.0 & 81.3 & 75.3 & 75.2 \\
    2018& 84.8 & 83.8 & 80.6 & 77.4 & 82.8 & 83.4 & 81.4 & 74.8 & 73.0 \\
    2019& 86.4 & 86.5 & 81.9 & 76.8 & 85.4 & 85.6 & 80.2 & 75.5 & 76.7 \\
    \hline
    AVG.& \textbf{84.2} & \textbf{84.2} & 79.9 & 77.0 & 82.3 & 82.8 & 80.9 & 75.1 & 75.1 \\
    \hline
    \end{tabular}
  }
\end{table}

\begin{itemize}
  \item As shown in Table~\ref{tab:acm_cls_result} and Table~\ref{tab:ieee_cls_result}, generally, the earlier calendar year $cy$ is, the lower F1 scores of models is. 
  The reason is that when the dataset is spilted by an ealier calendar year $cy$, more ``future'' Fellows elected after $cy$ will be labeled as non-Fellow. Thus, they are very similar to the Fellow-labeled examples and it is difficult to be classified.
  \item Cls-Fellow and Attention-RNN based model achieve better results, which are significantly better than other models. It demonstrates that time sequence model is more suitable for Fellow classification. The performance of attention-RNN based model and Cls-Fellow is very close, and on the IEEE classification dataset, former is slightly better. 
  \item GCN only use candidates’ co-author network information. The average F1 score of GCN can reach 79.9\% on IEEE dataset and 81.0\% in ACM dataset. It implies that the development of scholarly circle can represent candidates’ academic level to some extent.
\end{itemize}

For regression task, models are supposed to predict how many additional years it will take a candidate to become an IEEE/ACM Fellow. The Mean Absolute Error(MAE) on IEEE and ACM regression dataset is shown in Table~\ref{tab:acm_reg_result} and Table~\ref{tab:ieee_reg_result}. The observations are as follows:

\begin{table} [th]
  \newcommand{\tabincell}[2]{\begin{tabular}{@{}#1@{}}#2\end{tabular}}
\centering
\caption{The Mean Absolute Error (MAE) of Regression on ACM Regression Dataset}
\label{tab:acm_reg_result}
\resizebox{\columnwidth}{!}{
  \begin{tabular}{c|c|c|c|c|c|c|c|c|c}
  \hline
  &\tabincell{c}{Reg-\\Fellow}&\tabincell{c}{Attn-\\RNN}&\tabincell{c}{GCN-\\Only}&LR&Ridge&RF&DT&SVR&MLP\\
  \hline
  2009& 2.87 & 2.78 & 2.93 & 32.50 & 9.25 & 4.71 & 4.88 & 4.59 & 4.85\\
  2010& 2.30 & 2.27 & 2.54 & 14.76 & 4.96 & 4.21 & 4.41 & 4.10 & 2.92\\
  2011& 2.37 & 2.04 & 2.61 & 6.84 & 4.36 & 3.84 & 3.55 & 3.60 & 4.33\\
  2012& 1.86 & 1.71 & 2.02 & 5.19 & 4.08 & 3.30 & 3.44 & 3.19 & 2.92\\
  2013& 1.41 & 1.42 & 2.29 & 3.93 & 3.44 & 2.75 & 2.65 & 2.69 & 3.18\\
  2014& 1.23 & 1.44 & 1.79 & 4.19 & 3.30 & 2.22 & 2.13 & 2.28 & 2.44\\
  2015& 1.00 & 1.20 & 1.12 & 5.63 & 3.12 & 1.52 & 1.56 & 1.96 & 1.77\\
  2016& 0.67 & 0.88 & 0.98 & 3.90 & 2.91 & 1.02 & 1.02 & 1.67 & 1.10\\
  2017& 0.49 & 0.50 & 1.25 & 3.76 & 3.10 & 0.58 & 0.58 & 1.31 & 0.99\\
  2018& 0.24 & 0.07 & 0.35 & 3.86 & 3.04 & 0.31 & 0.03 & 1.15 & 0.46\\
  \hline
  AVG.& \textbf{1.44} & \textbf{1.43} & 1.79 & 8.46 & 4.16 & 2.45 & 2.43 & 2.65 & 2.50 \\
  \hline
  \end{tabular}
}
\end{table}

\begin{table} [th]
  \newcommand{\tabincell}[2]{\begin{tabular}{@{}#1@{}}#2\end{tabular}}
\centering
\caption{The Mean Absolute Error (MAE) of Regression on IEEE Regression Dataset}
\label{tab:ieee_reg_result}
\resizebox{\columnwidth}{!}{
  \begin{tabular}{c|c|c|c|c|c|c|c|c|c}
  \hline
  &\tabincell{c}{Reg-\\Fellow}&\tabincell{c}{Attn-\\RNN}&\tabincell{c}{GCN-\\Only}&LR&Ridge&RF&DT&SVR&MLP\\
  \hline
  2010& 2.47 & 2.67 & 4.74 & 3.89 & 3.77 & 3.84 & 3.81 & 4.42 & 4.81\\
  2011& 2.17 & 2.18 & 4.49 & 3.51 & 3.51 & 3.73 & 3.64 & 3.96 & 4.27\\
  2012& 1.86 & 1.94 & 3.74 & 2.96 & 3.07 & 3.25 & 3.40 & 3.54 & 3.49\\
  2013& 1.61 & 2.18 & 3.17 & 2.49 & 2.69 & 2.97 & 3.00 & 3.07 & 3.13\\
  2014& 1.33 & 1.65 & 2.59 & 2.25 & 2.34 & 2.48 & 2.52 & 2.68 & 2.79\\
  2015& 1.19 & 1.37 & 2.37 & 2.29 & 1.92 & 2.04 & 2.29 & 2.31 & 2.45\\
  2016& 0.96 & 1.17 & 1.57 & 1.47 & 1.57 & 1.71 & 1.75 & 1.94 & 1.74\\
  2017& 0.74 & 0.74 & 1.44 & 1.86 & 1.24 & 1.32 & 1.33 & 1.65 & 1.71\\
  2018& 0.49 & 0.49 & 1.10 & 1.36 & 1.09 & 0.80 & 0.79 & 1.21 & 0.75\\
  2019& 0.01 & 0.01 & 0.40 & 2.17 & 1.03 & 1.03 & 1.04 & 1.18 & 0.67\\
  \hline
  AVG.& \textbf{1.28} & \textbf{1.44} & 2.56 & 2.43 & 2.22 & 2.32 & 2.36 & 2.60 & 2.58 \\
  \hline
  \end{tabular}
}
\end{table}

\begin{itemize}
  \item It can be seen that Reg-Fellow model and attention-RNN based model achieves the lowest average MAE on the two regression datasets. Moreover, when the MAE of Reg-Fellow is less than 1, the calendar years are earlier than other models. Similar to the classification task, attention-RNN based model and our transformer encoder based model perform closely.
  \item For some traditional models, such as Ridge and SVM, although they handled the sequence data by taking the flattened data as input, they achieved worse performance.
  \item We noticed that the later calendar year $cy$ is, the smaller MAE gap between the GCN based model and Reg-Fellow model is. It may imply that the scholarly circle factor is more important in the late academic careers of candidates.
\end{itemize}

\subsection{Model Computation Time}
In the future, the Fellow prediction task and other career modeling tasks could be applied to some commercial online systems that require efficient calculation of large amounts of talent data. Therefore, computation time of prediction model are essential. According to the above experiments, we can find that self-attention based model (Cls-Fellow,  Reg-Fellow) and attention-RNN based models are significantly better than others. As show in Table~\ref{tab:time}, we compare their computation time on Fellow regression task (containing 4089 examples) and Fellow classification task (containing 56121 examples). In the view of computation time, two models are significantly different: the time cost of attention-RNN based models is usually about 40 times that of the transformer encoder based models.

\begin{table}
  \newcommand{\tabincell}[2]{\begin{tabular}{@{}#1@{}}#2\end{tabular}}
  \caption{Computation Time of Models}
  \label{tab:time}
  \begin{minipage}{\columnwidth}
    \begin{center}
      \begin{tabular}{c|c|c|c}
        \hline
        \textbf{Model} 
        & \textbf{\tabincell{c}{Batch\\Size}} 
        & \textbf{\tabincell{c}{Classification\\Computation Sec}} 
        & \textbf{\tabincell{c}{Regression\\Computation Sec}} \\
        \hline
        \multirow{3}*{\tabincell{c}{CLs-Fellow/\\Reg-Fellow}}  
        &  32 & 0.24 & 3.54\\
        &  64 & 0.17 & 2.62\\
        &  128 & 0.12 & 1.85\\
        \hline
        \multirow{3}*{\tabincell{c}{Attention-RNN\\based model}}  
        &  32 & 11.84 & 203.61\\
        &  64 & 6.12 & 102.12\\
        &  128 & 3.32 & 55.31\\
        \hline
      \end{tabular}
    \end{center}
  \end{minipage}
\end{table}

Through the above analysis, we can find that Cls-Fellow and Reg-Fellow model have excellent performance in terms of accuracy and efficiency. It shows that our proposed architecture is more robust and suitable for Fellow prediction than others.

\section{Discussion}
In this section, we try to discuss and verify some interesting results, such as the contribution of factors, the evolution of co-author networks, ``inequality'' (gender inequality),  and ``Good or Bad?'' (the reputation changes before and after being a Fellow). 

\subsection{The Contribution of Factors}
In Section~\ref{data}, we consider 8 kinds of factors with 36-dimensions to model the election of Fellows. Generally, some factors are essential to improve the competitiveness of candidates. Here we try to explore the contribution of different factors by visualizing the decision tree and analyzing the self-attention distribution.

\begin{figure*}
  \centering
  \includegraphics[width=1\columnwidth]{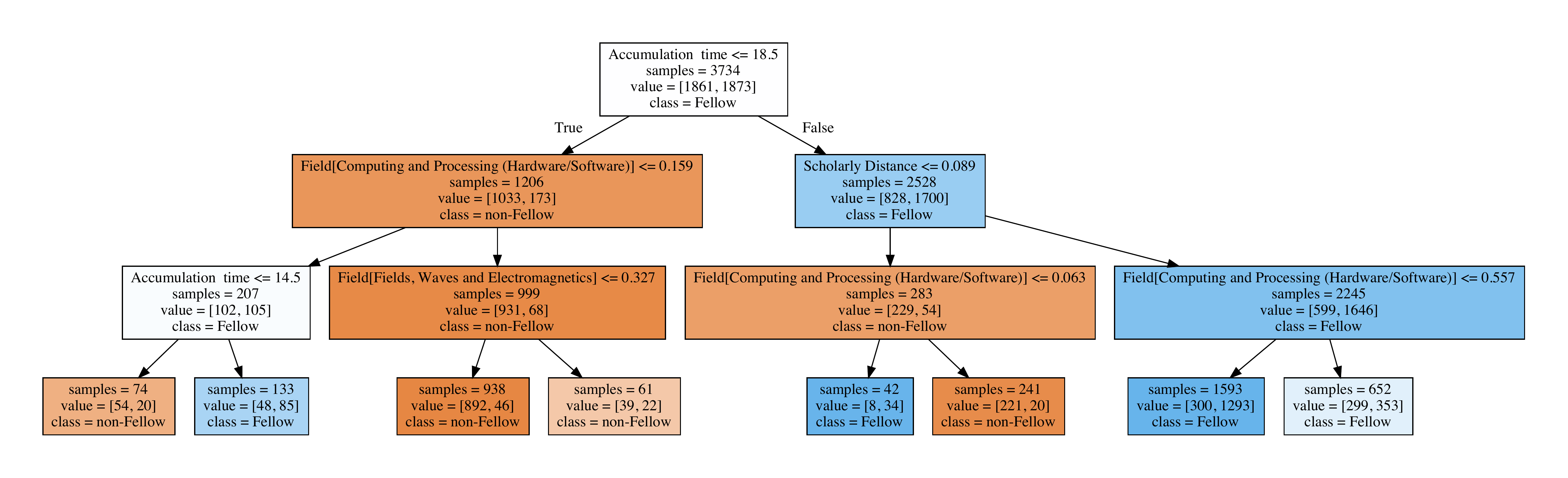}
  \caption{Visualization for the decision tree in IEEE Fellow classification.}
  \label{fig:ieee_tree}
\end{figure*}

\begin{figure*}
  \centering
  \includegraphics[width=1\columnwidth]{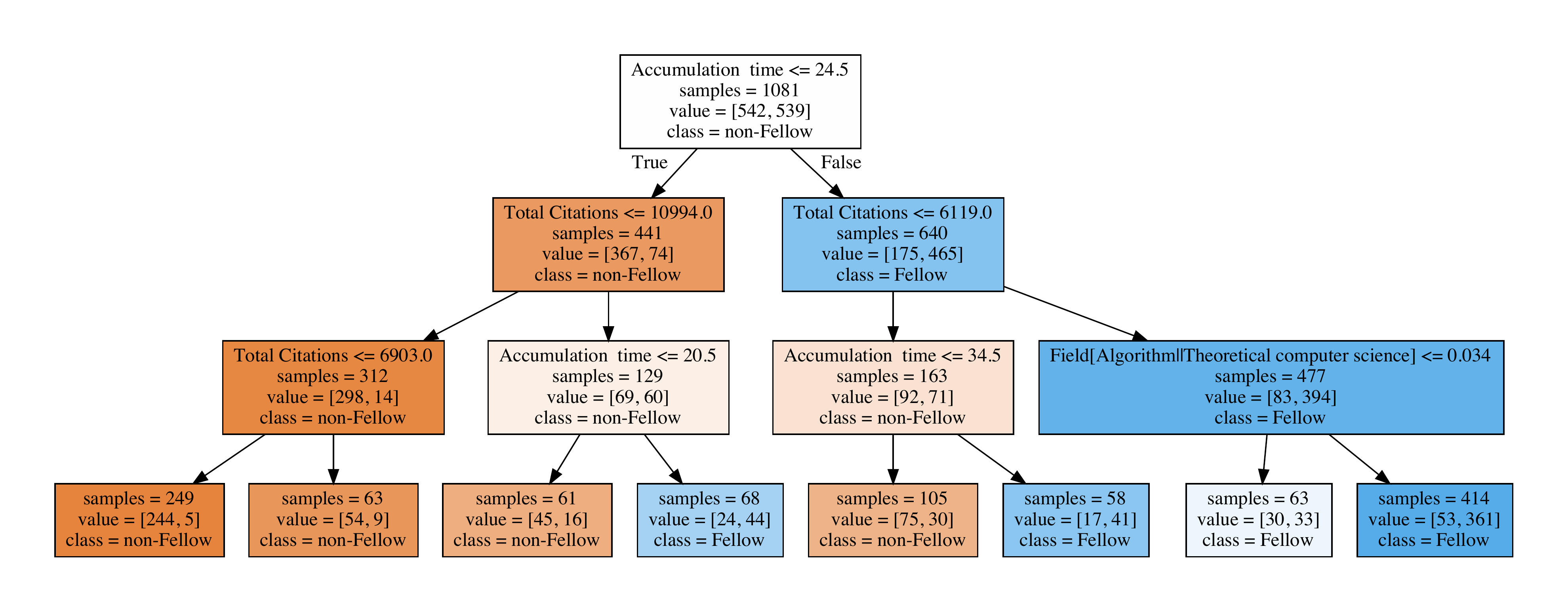}
  \caption{Visualization for the decision tree in ACM Fellow classification.}
  \label{fig:acm_tree}
\end{figure*}

Although the decision tree is not the best for classification and regression tasks, it is easy to be understood and interpreted by visualization. Here, we visualize the structures of the decision trees in IEEE Fellow classification and ACM Fellow classification, which are shown in Fig.~\ref{fig:ieee_tree} and Fig.~\ref{fig:acm_tree}. To facilitate visualization, these trees are trained on the calendar year $cy$ data (one year data) and also achieved acceptable performance (F1 score $> 80\%$ ). The conditions of each internal node in these trees are the key factors when distinguishing Fellow from non-Fellow. From Fig.~\ref{fig:ieee_tree} and Fig.~\ref{fig:acm_tree}, we can find that: 
\begin{itemize}
  \item In both in IEEE/ACM Fellow classification, the condition of accumulation time is always at the root. It means that, in the view of decision tree, the accumulation time is the most critical factor. ACM candidates with more than 24 years of accumulation and IEEE candidates with more than 18 years of accumulation are generally more likely to be elected as Fellows. 
  \item Total citations is also worthy of attention. In the ACM decision tree, as shown in Fig.~\ref{fig:acm_tree}, the total number of citations and accumulation time are combined as the rule for ACM fellow classification: Candidates with more than 24 years of academic output and more than 6,119 citations are more likely to be elected as Fellows and the rising stars (candidates with shorter accumulation time) are able to be very competitive due to a higher number of citations. It means that academic accumulation and influence are comprehensively considered in Fellow elections. We also notice that candidates with less than 6903 citations are usually difficult to be elected as ACM Fellows. 
  \item Candidates' embedding scores in some research fields, such as Computing and Processing (Hardware/Software) in IEEE, are used as conditions, especially in IEEE decision tree. It shows that the difficulty of becoming a Fellow in various research fields may be different. As we all know, compared with ACM, IEEE's research fields are broader and more diverse, which may be a factor that the field embedding scores are usually used as conditions in IEEE decision tree.
  \item Scholarly distance plays an import role in IEEE Fellow election, which indicate candidates need to strengthen academic cooperation and develop their academic social networks.
\end{itemize}

\subsection{The evolution of co-author networks}

With the accumulation of candidates' publications, their co-author networks are also growing. The close cooperation between candidate and elected Fellows not only makes it easier for candidate to find endorsers during the Fellow election, but also may imply that the candidate’s research work is excellent. In Section~\ref{data}, we introduce \textbf{Scholarly  distance} to measure the distance between a candidate and existing Fellows and calculate the co-author network embedding for \textbf{Scholarly circle} representation. Although, scholarly distance and scholarly cricle play important roles in Fellow regression task and Fellow classification task, they are not intuitive for human understanding and reference. Here, we introduce the number of collaborations with Fellows ($N_{collab}$) and the number of Fellow co-authors ($N_{neighbor}$) to indicate the distance between candidates and existing Fellows. we explore the difference between IEEE Fellows, ACM Fellows and non-Fellows by analyzing the evolution of their $N_{collab}$ and $N_{neighbor}$ over the accumulation time. 

In order to ensure the comparability of the data, the scholars used for co-author network analysis and comparison should be the same generation, which means that their publication year range should be the same. According to the previous analysis, it usually takes 20-25 years for a talented scientist to be elected as an ACM/IEEE Fellow. Thus, we select scholars from IEEE Fellow dataset, ACM Fellow dataset and non-Fellow dataset based on the publication years and elected years, and the details are as follows:
\begin{itemize}
  \item For \textbf{IEEE/ACM Fellow}, we selecte IEEE/ACM Fellows who published their first publications between 1995 and 2000 and were elected as IEEE Fellow or ACM Fellow after 2015.
  \item For \textbf{non-Fellow}, we selecte scholars from the non-Fellow dataset which contains ACM distinguished scholars and Aminer highly cited scholars. These schoalrs also published their first publications between 1995 and 2000, but they have not yet been elected as IEEE/ACM Fellow.
\end{itemize}

\begin{figure*}[ht]
    \begin{subfigure}{0.32\textwidth}
      \centering
      \captionsetup{justification=centering}
      \includegraphics[width=1\linewidth]{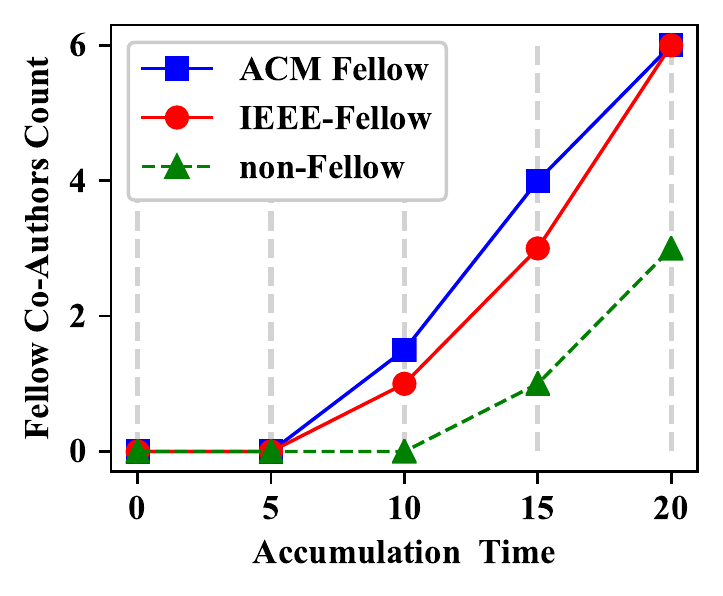}
      \caption{The Median of Fellow Co-Authors within 1 Hop}
      
    \end{subfigure}
    \begin{subfigure}{.32\textwidth}
      \centering
      \captionsetup{justification=centering}
      \includegraphics[width=1\linewidth]{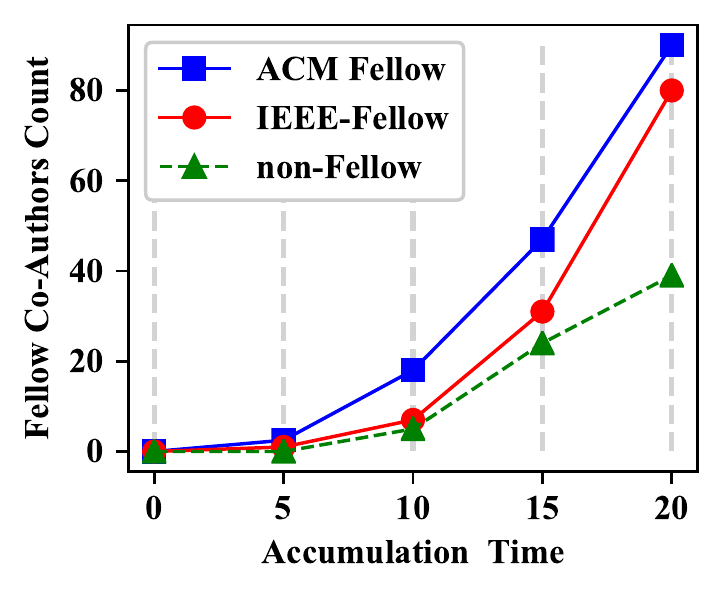}
      \caption{The Median of Fellow Co-Authors within 2 Hop}
    \end{subfigure}
    \begin{subfigure}{.32\textwidth}
      \centering
      \captionsetup{justification=centering}
      \includegraphics[width=1\linewidth]{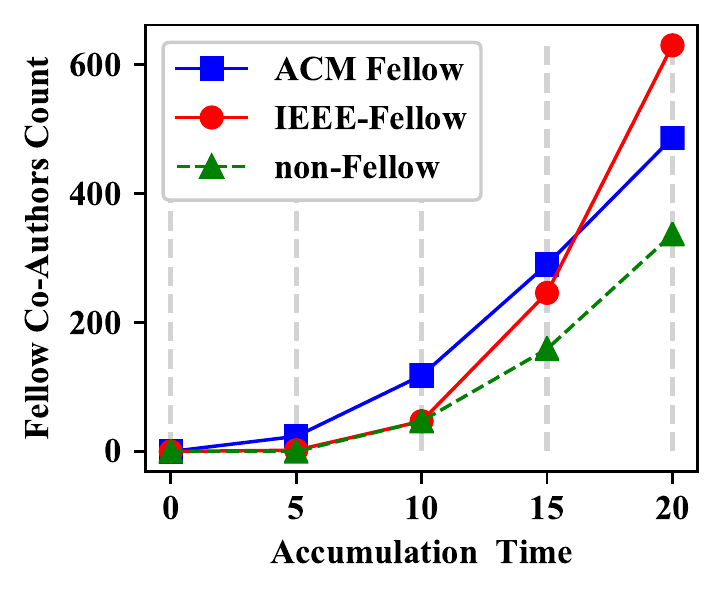}
      \caption{The Median of Fellow Co-Authors within 3 Hop}
    \end{subfigure}
  
  
    \begin{subfigure}{0.32\textwidth}
      \centering
      \captionsetup{justification=centering}
      \includegraphics[width=1\linewidth]{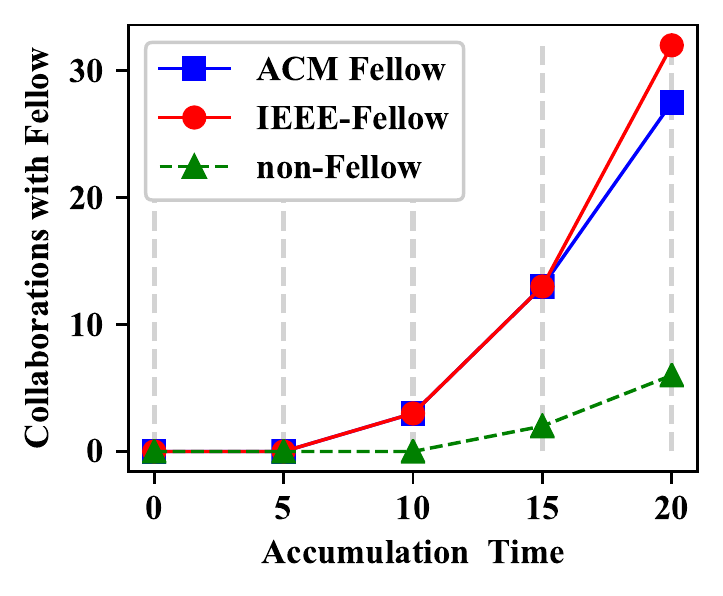}
      \caption{The Median of Collaborations with Fellow within 1 Hop}
    \end{subfigure}
    \begin{subfigure}{.32\textwidth}
      \centering
      \captionsetup{justification=centering}
      \includegraphics[width=1\linewidth]{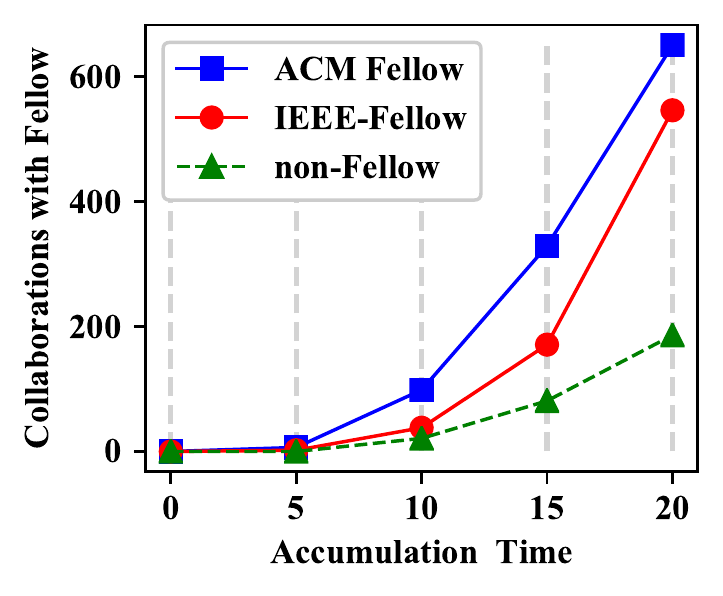}  
      \caption{The Median of Collaborations with Fellow within 2 Hop}
    \end{subfigure}
    \begin{subfigure}{.32\textwidth}
      \centering
      \captionsetup{justification=centering}
      \includegraphics[width=1\linewidth]{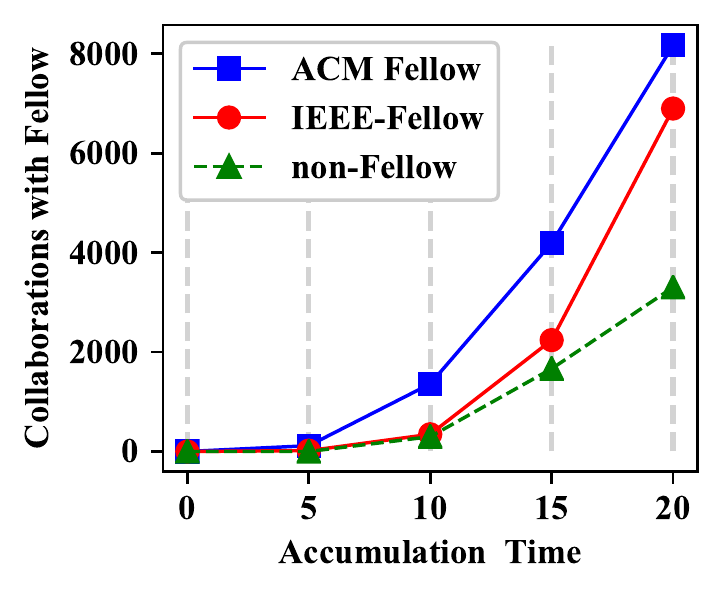}  
      \caption{The Median of Collaborations with Fellow within 3 Hop}
    \end{subfigure}
  \caption{The evolution of Fellow co-authors and Fellow-related collaborations in co-author networks}
  \label{fig:co-network-line}
  \end{figure*}

We define three scopes, namely 1-hop, 2-hop and 3-hop, which are used to limit the scope when calculating each candidate's co-author network. 1-hop only includes candidate’s direct co-authors. The scholars in 1-hop and their direct co-authors are included in 2-hop. 3-hop is the widest scope which covers 1-hop and 2-hop and includes direct co-authors of scholars in them. For each scholar we selected from IEEE Fellow, ACM Fellow and non-Fellow dataset, we calculate their $N_{collab}$ and $N_{neighbor}$ over the accumulation time. The evolution of them over the accumulation time in three scopes are shown in Fig.~\ref{fig:co-network-line}. From the Fig.~\ref{fig:co-network-line}, we have the following findings:
\begin{itemize}
  \item Both IEEE/ACM Fellows and non-Fellows can promote their close collaboration with existed Fellows. 
  \item Although non-Fellows are outstanding, compared with IEEE or ACM Fellows, their co-author networks with existed Fellows grow slowly in terms of speed and value.
  \item The difference between non-Fellows and IEEE/ACM Fellows in the first 5 years is not obvious, and the difference starts to show up in 5-10 years. In the early 10 years, IEEE/ACM Fellows usually have had the direct cooperation with existed Fellows. Previous studies~\citep{li2019early} find that junior researchers who coauthor work with top scientists enjoy a persistent competitive advantage throughout the rest of their careers. It implies that cooperation with existed Fellows in candidates' early careers play a key role for their academic development.
  \item In 2-hop scope and 3-hop scope, non-Fellows and IEEE Fellows are relatively close in the first 10 years. However, as shown in sub-figure $(a)$ and sub-figure $(d)$ of Fig.~\ref{fig:co-network-line}, in the first 10 years, IEEE Fellows usually have had direct cooperation with existed Fellows, but non-Fellows have not. It implies that an important challenge of becoming a Fellow is how to transform Fellows who have indirect cooperation with candidates into their direct collaborators. 
  \item As show in Fig.~\ref{fig:co-network-line}, the $N_{collab}$ and $N_{neighbor}$ of ACM Fellows are usually more than that of IEEE Fellows. It may be caused by the fact that the research fields of ACM are more concentrated than those in IEEE, which is more conducive to cooperation.
\end{itemize}

\subsection{The Inequality of Gender}
\label{sec:female}

To check the inequality of gender, we divide the Fellows into two groups according to their genders, and we calculate the $\bar{\alpha}$ values and standard deviation $\sigma(N'(t))$ of the average productivity $<N'(t)>$ between male Fellows and female ones. Here $<N'(t)>$  is the average properties of $N_i(t)$ for all scientists in one group by defining the normalized average trajectory as follows:
\begin{equation}
\label{equ:trajectory}
<N'(t)>=\frac{1}{I}\sum_{j=1}^{j=I}\frac{N_j(t)}{\overline{n}_j},
\end{equation}
\noindent where $<N'(t)>\sim t^{\bar{\alpha}}$, $\overline{n}_j$ is the average annual production of scientist $i$ and $\sigma(N'(t))$ is the standard deviation of $N'(t)$.

\begin{figure}
\centering
\includegraphics[width=0.6\columnwidth]{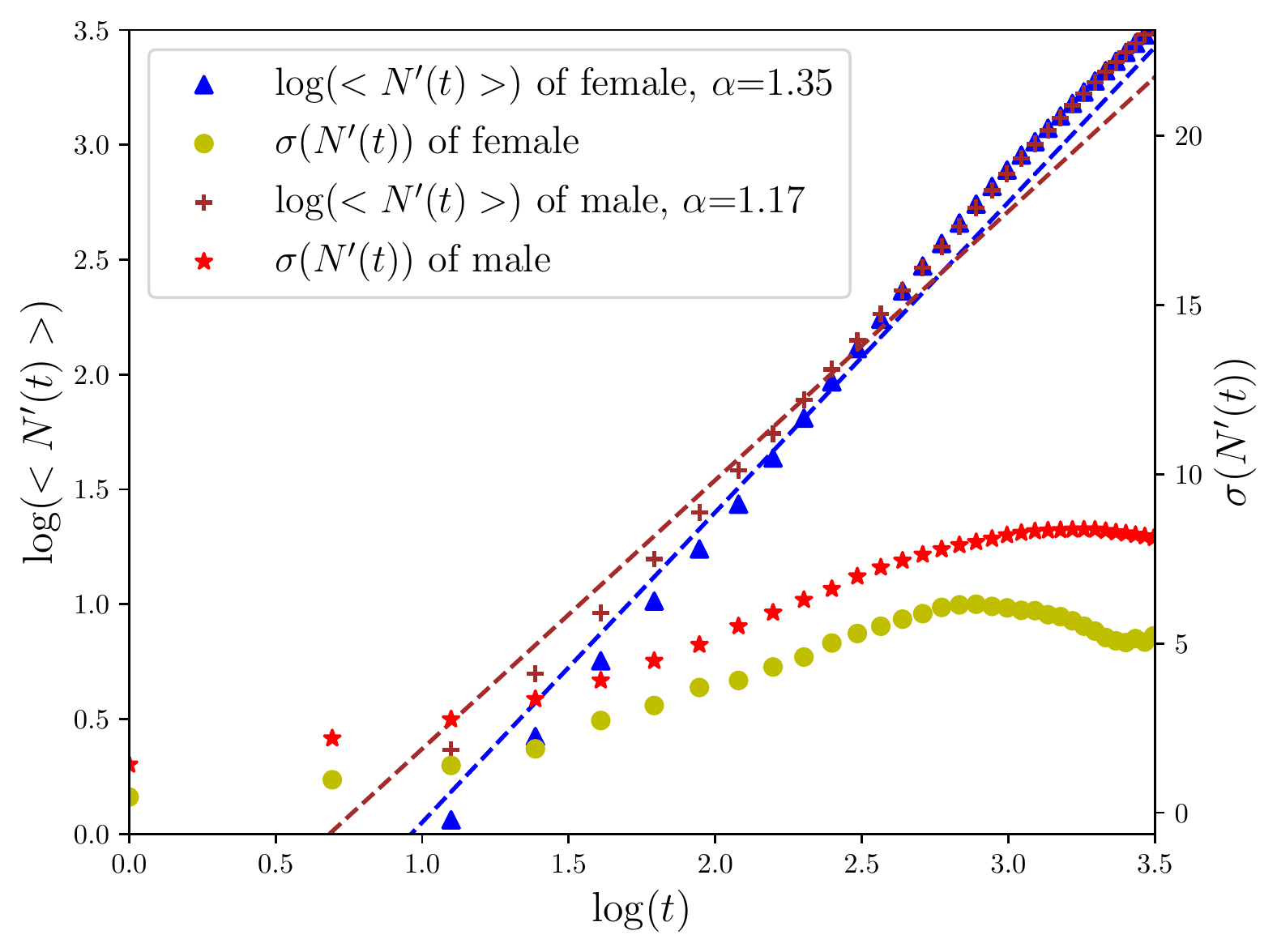}
\caption{The scholarly productivity $<N'(t)>$ and $\sigma(N'(t))$ between male and female Fellows. Note that the left axes are in log scale (best viewed in color).}
\label{fig:gender-pub-alpha}
\end{figure}

The two lines in Fig.~\ref{fig:gender-pub-alpha} show the relationship between $<N'(t)>$ and $t$ (log scale) for male Fellows and female Fellows, respectively. We can observe that female Fellows are significantly different to male Fellows in terms of $\bar \alpha$ value ($p=2 \times 10^{-5}<0.001$, Mann-Whitney Test). It indicates that female Fellows need to put in more effort than male Fellows if they are elected as Fellows.
Moreover, the two curves show the trends between $\sigma(N'(t))$ and $t$. Generally,  a broad peak is a likely signature of career shocks that can significantly alter the career trajectory~\citep{Petersen2012}. In the early years of their careers, male Fellows have higher academic productivity than female Fellows. However, female candidates are growing faster.

Based on our data, the above results indicate that the inequality of gender does exist in the Fellow selection, like the inequality in faculty hiring network, grant proposal, etc. 

\subsection{Good or Bad: The Reputation Change Before and After Being a Fellow}
\begin{figure}
\centering
\includegraphics[width=0.6\columnwidth]{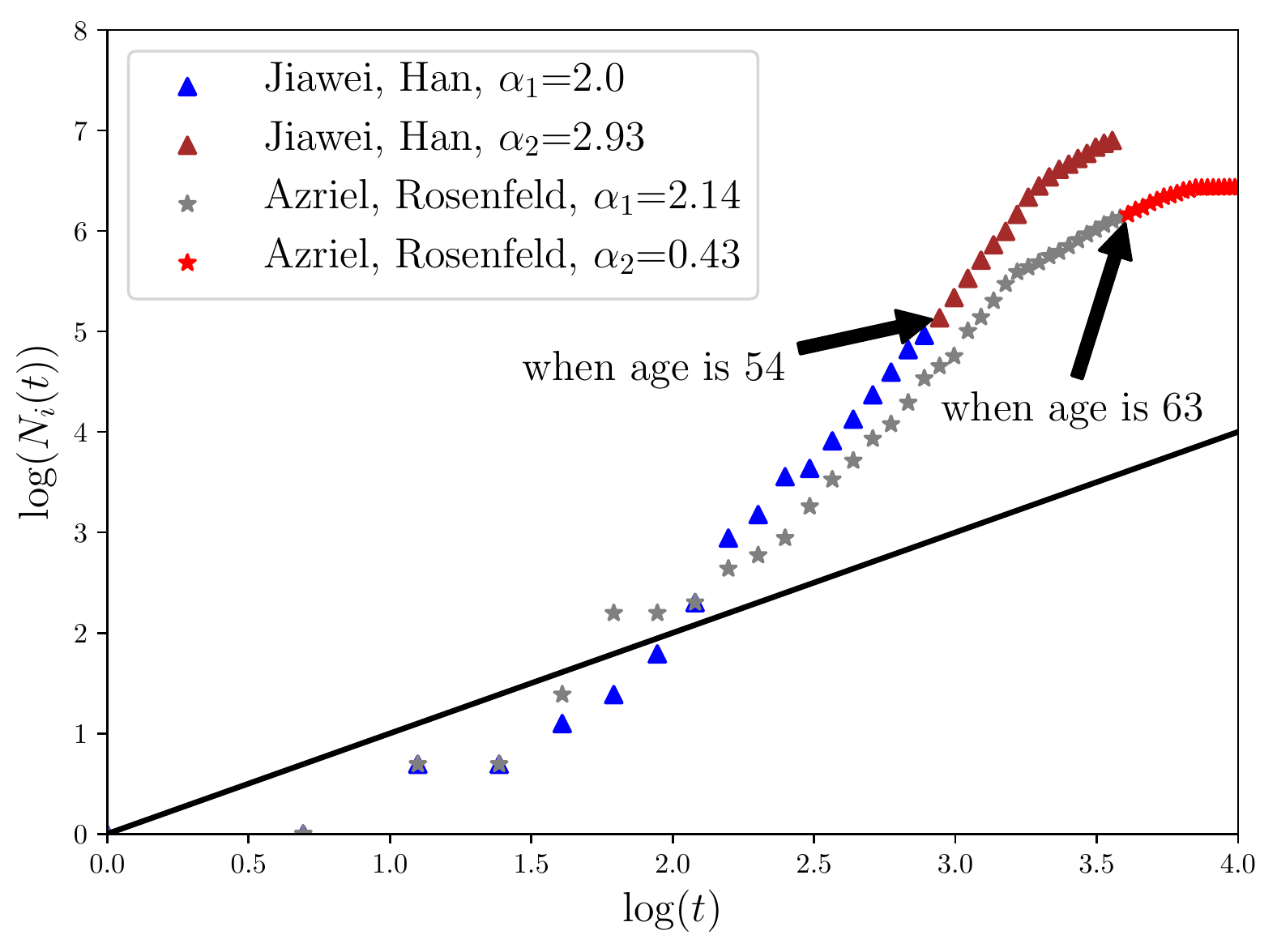}
\caption{The Reputation Change before and after Fellow of 2 examples (best viewed in color).}
\label{fig:2-alpha}
\end{figure}

Another phenomenon we are interested in is whether the Fellow title can bring about a good result for a scientist's reputation. Here we monitor
 the change trends of his/her productivity factor $\alpha$ and average citations per paper, thus we define $ \alpha_{i,1} $ as the $\alpha$ value of publications of scientist $i$  before his/her Fellow nomination, and $ \alpha_{i,2} $ as the $\alpha$ value after he/she is selected as a Fellow.  Fig.~\ref{fig:2-alpha} shows the change of two Fellows, it reveals that the productivity of Fellow Han Jiawei increases dramatically, while Fellow Rosenfeld Azriel decreases dramatically. The age when elected may be the reason for this difference. We find that the first quartile is 1.28, the median is 1.53, and the third quartile is 1.78 before being a Fellow. After being a Fellow, the first quartile is 0.60, the median is 1.11, and the third quartile is 1.62. Then we define the difference of $\alpha$ before and after being a Fellow as $\alpha_{diff} = \alpha_{i,2} - \alpha_{i,1}$. And we analyze the ratio between Fellows with $\alpha_{diff}>0$ and those with $\alpha_{diff}<0$, we find that the ratio is about $1:2.35$. Moreover, we also calculate the change of citations per paper of a scientist before and after his/her Fellow nomination. We find the citations per paper are decreased from 91.8 to 31.6 and the ratio between increase and decrease on citations per paper is $1:6.8$. The results demonstrate that a scientist before being a Fellow has achieved high-impact publications, i.e., his/her high-cited papers are published before he/she is selected as the Fellow. That is, being a Fellow could not bring about the improvement of citations for a scientist. This discovery are similar to that in Ref. ~\citep{Brogaard2018}, which indicates that researchers have reduced high-risk, highly innovative research after receiving tenure.

\section{Conclusion and Future Work}
In this paper, we tried to explore the key factors of Fellow selection and proposed two self-attention based model to classify Fellows or non-Fellows and predict how many years it takes a talented candidate to be elected as a Fellow. Moreover, we analyzed the factors of the Fellow nomination qualitatively and defined a scholarly distance to measure the co-author network between a candidate and existing Fellows. We also discover some interesting phenomena from the Fellow datasets, such as the evolution of co-author networks between candidates and Fellows, inequality of gender and the change of reputation with/without the Fellow title. It is worth noting that the conclusions and observations are reached based on the current datasets, which maybe are not inconsistent with the practical process of Fellow election. However, we believe that the talented researchers and Fellow Committees can benefit from our findings for Fellow election and nomination.
Despite we only focus on the data of IEEE/ACM Fellows in this paper, the relevant research ideas and models can be extended to other data. In the future, when the data is available, we are pleased to explor on other dataset.

\section{Appendix}

\subsection{Data Statistics and Analysis}
\label{data_analysis}
\subsubsection{Accumulation Time}
\begin{figure}
\centering
\includegraphics[width=0.9\columnwidth]{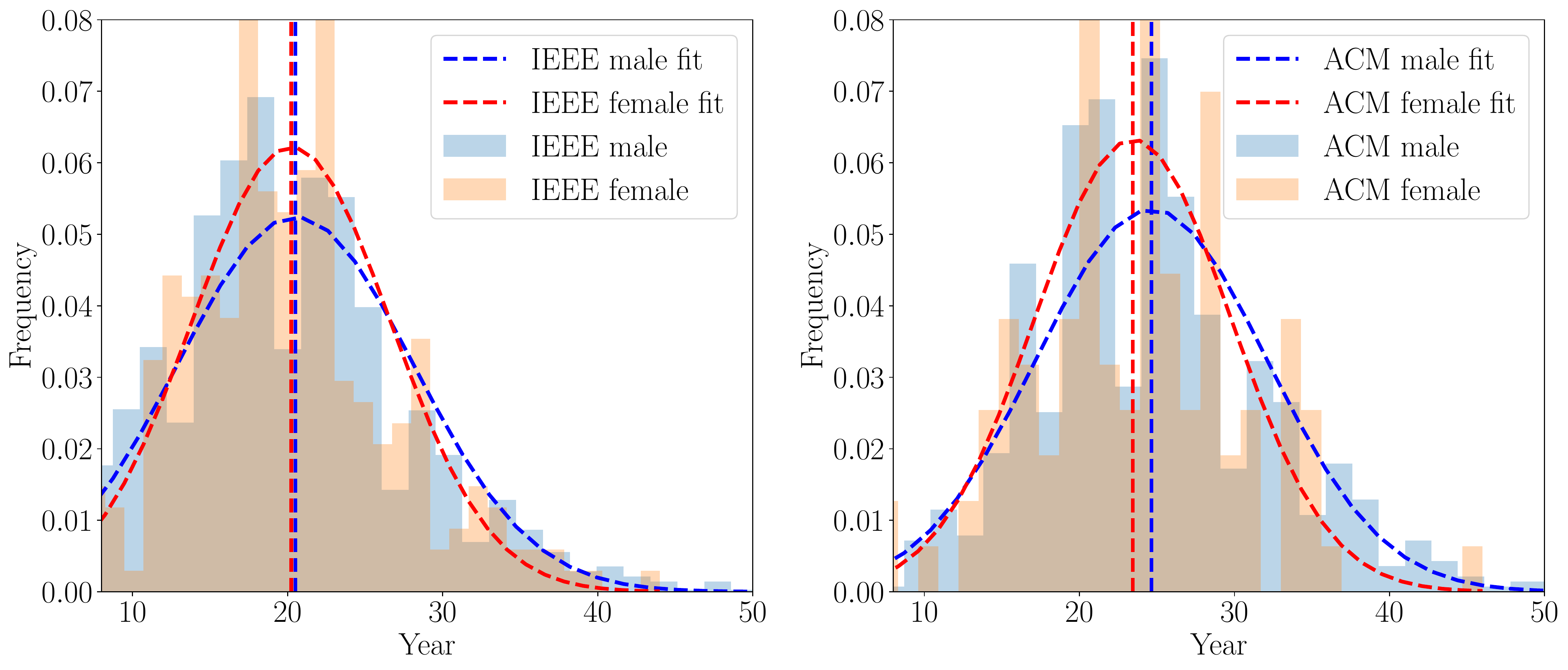}
\caption{The distributions of the years to be selected as ACM and IEEE Fellows (best viewed in color).}
\label{fig:wait-year}
\end{figure}

We compute the distribution of accumulation time $t$ for all Fellows in our datasets, as shown in Fig.~\ref{fig:wait-year}. We can observe that the distributions of $t$ of IEEE and ACM Fellows both obey normal distribution appropriately. From a gender perspective, we can find that: (1) The $\sigma$ of females are lower than that of males ($\sigma_{male}^{IEEE}=7.61,\sigma_{female}^{IEEE}=4.61,\sigma_{male}^{ACM}=7.45,\sigma_{male}^{ACM}=6.31$). (2) IEEE male Fellows and IEEE female Fellows have the similar means $\mu$ ($\mu_{male}^{IEEE}=20.50,\mu_{female}^{IEEE}=20.23$). However, in ACM, female Fellows have a slightly smaller mean $\mu$ than male Fellows ($\mu_{male}^{ACM}=24.65,\mu_{female}^{ACM}=23.45$).
We also use the Mann-Whitney Test~\citep{nadim2008} to explore the difference of accumulation time between male Fellows and female Fellows. It shows that there is no significant difference of years between male Fellows and female Fellows ($p^{IEEE}=0.48>0.05, p^{ACM}=0.12>0.05$, Mann-Whitney Test~\citep{nadim2008}).
Moreover, we also observe that ACM Fellows spend more 3 to 4 years than IEEE Fellows on average. The above results demonstrate that time plays a more important role in Fellow election for different organizations, rather than different genders. In our proposed model, for each candidate, we consider the accumulation years from his/her first publication year to a given year as one factor. 

\subsubsection{Scholarly Productivity}
\begin{figure}
\centering
\includegraphics[width=0.6\columnwidth]{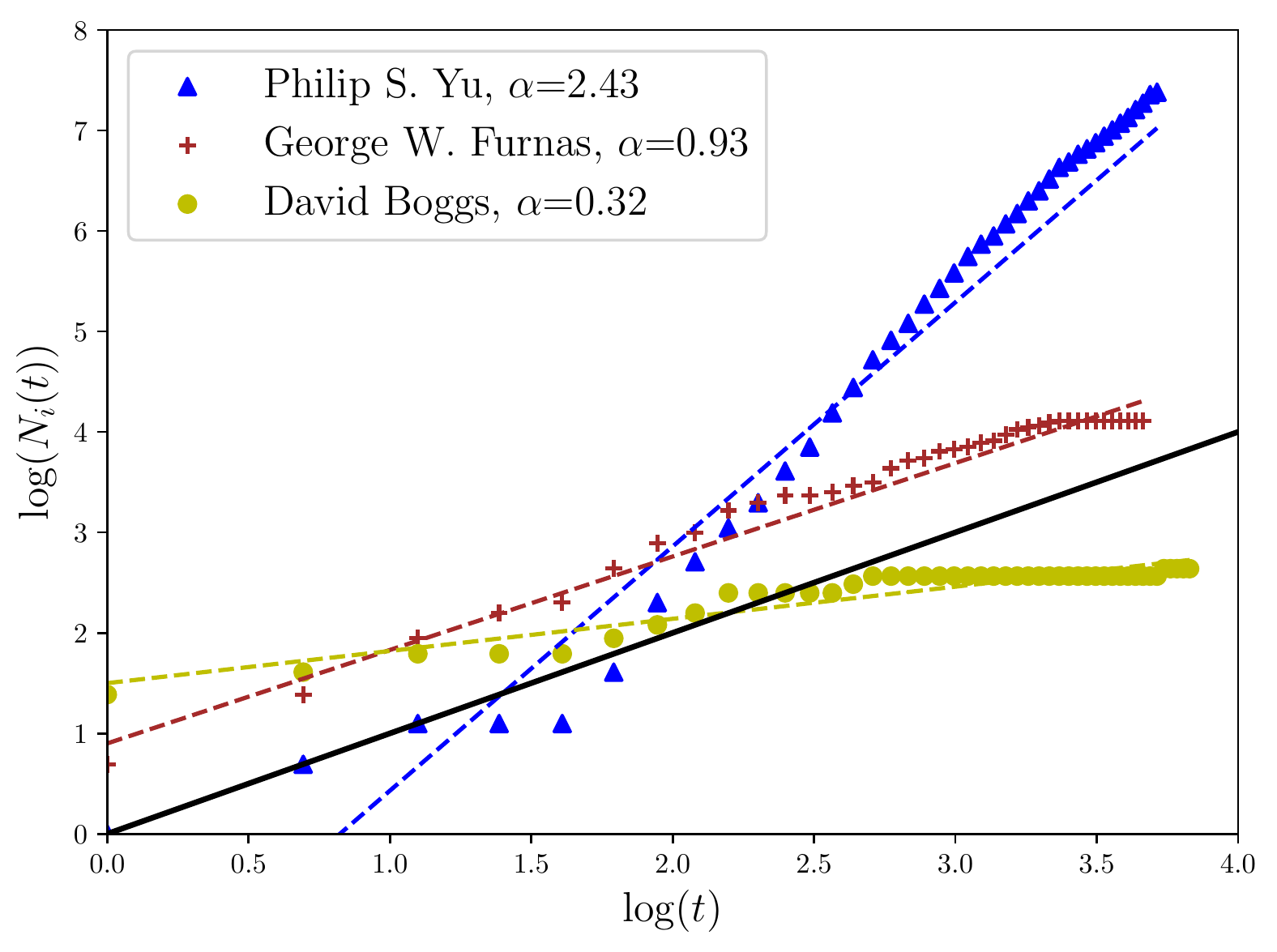}
\caption{The scholarly productivity measured by $\alpha$ of three Fellows (best viewed in color).}
\label{fig:pub-alpha}
\end{figure}

As described in Section~\ref{data}, we calculate $\alpha$ to quantify the scholarly productivity.
Fig.~\ref{fig:pub-alpha} shows three Fellows with different $\alpha$ values. It is obvious that the publications of Prof. Philip S. Yu increase much faster than Prof. David Boggs and George W. Furnas and his scholarly productivity increases almost twofold during his career, while the increase is slower for Prof. Kahan, William. We can divide the Fellows into ultrahigh-productivity ($\alpha\geq2$), high-productivity ($1< \alpha< 2$), and modest-productivity ($\alpha \leq 1$) ones manually according to the value of $\alpha$, and the ratio among the three categories is $2.7:2.5:1$. We can find that 84\% Fellows in our datasets have a steady increase in their productivity with time $t$. 

\subsubsection{Visualization}
\begin{figure}
  \centering
  \includegraphics[width=0.8\columnwidth]{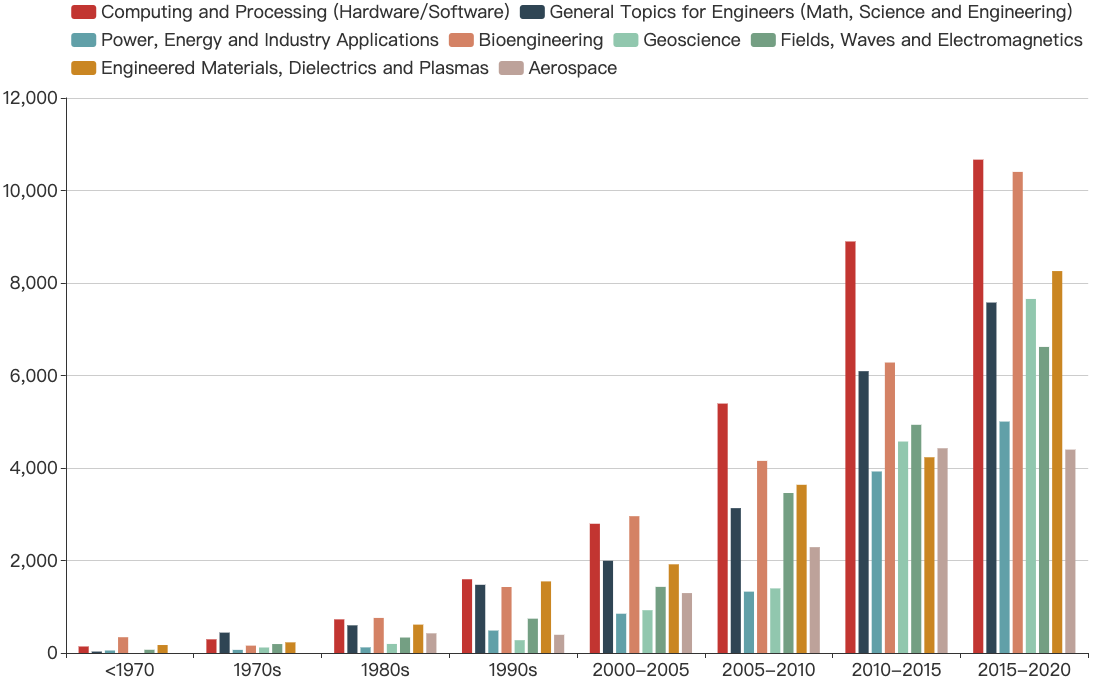}
  \caption{The third quartile of citations when IEEE Fellows were elected.}
  \label{fig:fellow_75th_percentile}
\end{figure}

We want to explore the differences in the difficulty of becoming a Fellow in different research fields. To this end, for Fellows in each research field, we calculate the average and third quartile of their academic features. Considering that the criteria of ACM/IEEE Fellow election are evolving over time, the data we calculate are grouped by time. The academic features of Fellows are the factors we consider in Section~\ref{data}, including $h-index$, $\alpha$ of citations, etc., which can be used as the goals for candidates. The Fig.~\ref{fig:fellow_75th_percentile} shows the difference in citations of IEEE Fellows from various research fields and reveals the evolution of citations over time. As can be seen from the Fig.~\ref{fig:fellow_75th_percentile}, the number of citations of candidates from aerospace field is generally less than candidates from computing and processing (Hardware/Software) field , when they were both elected as IEEE Fellow in the same year. We also plot histograms about other academic features for IEEE/ACM Fellows in the website \footnote{https://fellow.dawn-story.cn/}. We believe that these histograms can help people understand the trend of the IEEE/ACM Fellow election and realize the gap between themselves and IEEE/ACM Fellows.

\subsection{Experiment Data preparation}
\label{exp_prepar}
\subsubsection{Classification Data}
In classification task, we adopt the examples in ACM/IEEE Fellow datasets as positive class, and the examples in non-Fellow datasets as negative class. Specifically, We introduce a calendar year $cy$ as variable. 

For each classification dataset, in a given calendar year $cy$, Fellows who were elected as IEEE/ACM Fellows before $cy$ will be considered as the positive examples (labeled Fellow), then the rest will be considered as the negative examples (labeled non-Fellow). After training and testing classification models on the datasets splitted by different calendar years, we can evaluate their predictive performance and robustness.

For each candidate $i$ in classification datasets, we used a matrix concatenated by vectors in $X_i$ ($X_i$ is shown in Eq.~\ref{equ:fellow_represent}) to represent his/her academic trajectory from the year $sy_i$ to the given calendar year $cy$.

\subsubsection{Regression Data}
\begin{figure}
  \centering
  \includegraphics[width=0.5\columnwidth]{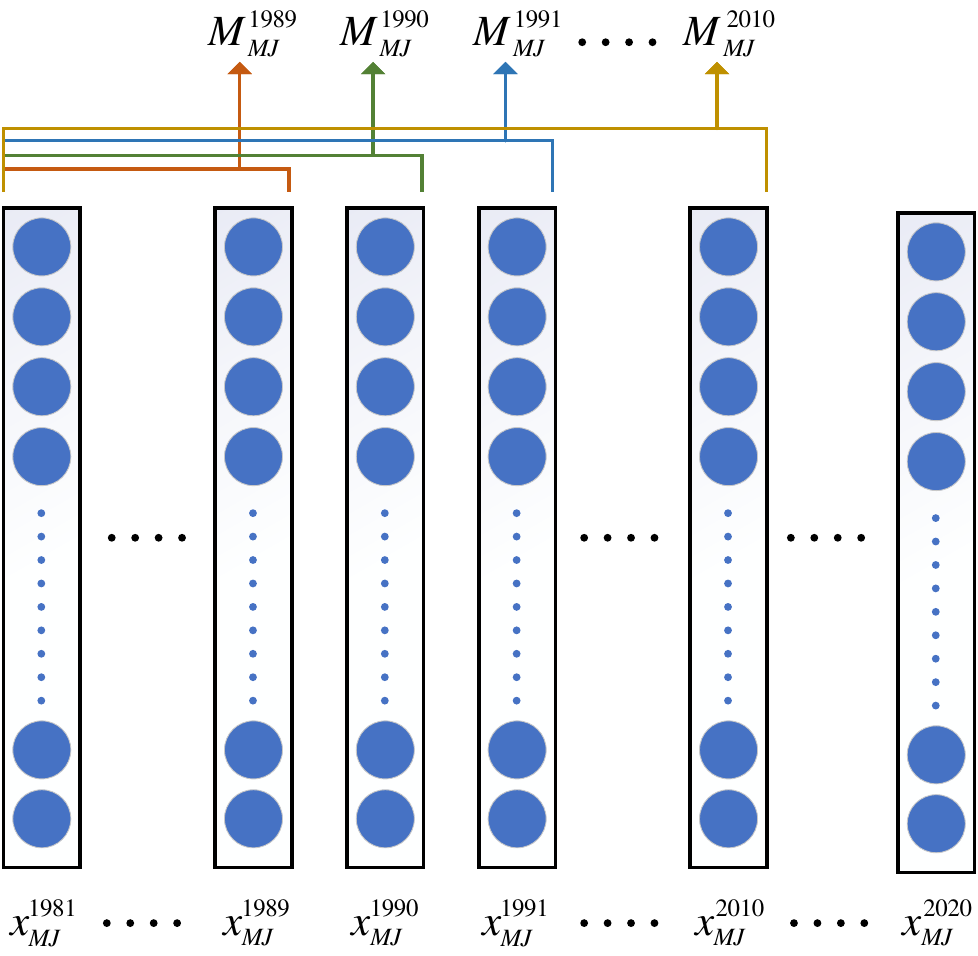}
  \caption{Graphical illustration of input matrices from a Fellow for Reg-Fellow.}
  \label{fig:reg_data}
\end{figure}

For Fellow regression task, the input is a matrix $M_i^{y_j}$ representing a Fellow $i$'s academic trajectory from the year $sy_i$ when he/she first published his/her paper to a certain year $y_j$. $ey_i$ represents the year when he/she was elected as Fellow. And $ey_i-y_j$, as the target of regression, represents how many additional years the candidate will take to be named as Fellow. Each row of matrix $M_{i}^{y_j}$ is Fellow $i$'s annual factor vector. In our Fellow dataset, candidates, such as Robert G. Gallager and Alan Laub, can be selected as Fellows as early as the eighth years after they published first publications. Thus, we define that $y_i$ is between the 8th year after $sy_i$ and $ey_i$. Due to the value of $y_j$, Fellow $i$ 's time series vectors $X_i$ can be selected and concatenated into $ey_i-sy_i-8$ matrices as the input of regression models, as shown in Eq.~\ref{equ:fellow_represent} and Eqs. \ref{equ:reg_inputs}-\ref{equ:reg_fellow_data_matrices}.
\begin{equation}
  \label{equ:reg_inputs}
  \mbox{Reg-Fellow's input}= \bigcup_{i=1}^{N}\mathbf{M}_i,
\end{equation}
\begin{equation}
  \label{equ:reg_one_fellow_set}
  \mathbf{M}_{i}=\{M_{i}^{sy_i+8},\cdots, M_{i}^{y_j} , \cdots, M_{i}^{ey_i}\}, sy_i + 8 \leq y_j \leq ey_i,
\end{equation}
\begin{equation}
  \label{equ:reg_fellow_data_matrices}
  M_{i}^{y_j} = concat(\mathbf{x}_{i}^{sy_i} , \cdots, \mathbf{x}_{i}^{y_j}),
\end{equation}
\noindent where the $N$ is the total Number of IEEE/ACM Fellows we collected. The $\mathbf{M}_i$ is a matrix set which contains all matrices built from an IEEE/ACM Fellow $i$'s time series vectors $X_i$. For example, “Michael I. Jordan” published his first paper in 1981 and he was elected as an ACM Fellow in 2010, so his input matrices are shown in Figure~\ref{fig:reg_data}. 

Finally, for Fellow regression task, there are 57,979 matrices from IEEE Fellows and 11,095 matrices from ACM Fellows. We called the collections of all IEEE/ACM Fellows' academic trajectories as IEEE regression dataset and ACM regression dataset, respectively. 

Similar to the classification task, we introduced calendar year $cy$, a variable, to split dataset into training set and test set. For each dataset, in a given calendar year $cy$, for example if we are in 2017, all the Fellows in and before 2017 will be considered as the training set, then the rest Fellows who are selected after 2017 will be considered as the testing set.

\subsection{Hyper-parameter Settings}
\label{param_setting}
\begin{figure*}[ht]
    \begin{subfigure}{0.49\textwidth}
      \centering
      \captionsetup{justification=centering}
      \includegraphics[width=1\linewidth]{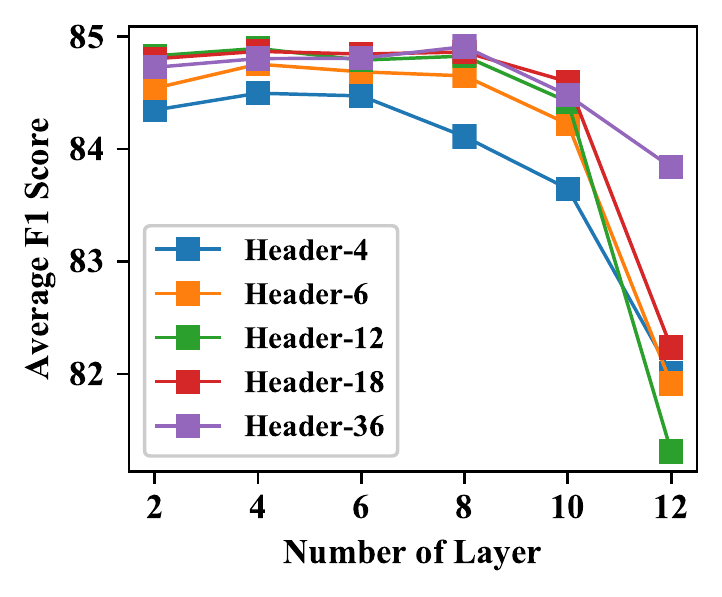}
      \caption{Fellow Classification Task}
      
    \end{subfigure}
    \begin{subfigure}{.49\textwidth}
      \centering
      \captionsetup{justification=centering} 
      \includegraphics[width=1\linewidth]{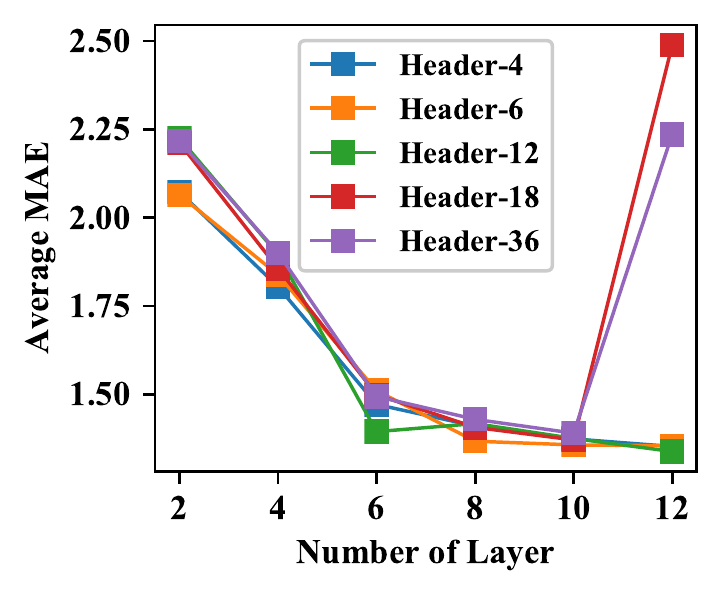}
      \caption{Fellow Regression Task}
    \end{subfigure}
  \caption{The Performance of Our Models With Different Parameter Settings}
  \label{fig:param-analysis}
\end{figure*}

In our proposed Cls-Fellow model and Reg-Fellow model, the number of encode layers and the number of heads in each layer affect the predictive performance. Therefore, we conducted two experiments on regression task and classification task for these key parameters. As shown in Fig.~\ref{fig:param-analysis}, our models often achieve better performance when the number of encode layers is around 8. When the number of encode layers are fixed, we can compare the performance of different number of head. In Fellow classification task, Cls-Fellow model with 36 heads performs best. And in Fellow regression task, it is better to set the number of heads to 6 for Reg-Fellow model.

 During the training, the initial learning rate is 0.001 and the batch size is 128. Adam optimization method~\citep{KingmaB15} is used to find the best weights in 20 epochs. These models are implemented with Tensorflow 2.0 in Python.

 \subsection{Implement Details of Baseline Methods}
 \label{baseline_detail}

 \subsubsection{Classification Baselines}
 \begin{itemize}
  \item \textbf{SVM:} $\epsilon$-Support Vector machine, its penalty parameter $C$ is $10$.
  \item \textbf{LR:} LR fits a linear model with coefficients to minimize the residual sum of squares between the observed responses in the dataset, and the predicted responses.
  \item \textbf{Ridge:} Ridge regression improves LR by imposing a penalty on the size of coefficients, and the parameter $\alpha$ in Ridge is set as 0.1 after tuning.
  \item \textbf{Decision Tree:} We used Gini impurity as the split criteria. After tuning, the parameters are as following: the max depth is 3 and the minimum number of samples is set to 32.
  \item \textbf{Random Forest:} A random forest is a meta estimator that fits a number of decision tree classifiers on various sub-samples of the dataset and uses averaging to improve the predictive accuracy and control over-fitting. After tuning, the number of trees in the forest was set as 32 and other parameters are the same as the decision tree.
  \item \textbf{GCN:} We only use candidates’ co-author network data as input data. We utilize two GCN layers and the outputs of second GCN layer are aggregated to a 36-dimensional vector by max pooling. Finally, the sequence of graph embedding are processed by 8 transformer encode layers with 12 heads. 
  \item \textbf{MLP:} The hidden layer sizes are 64, 32 and 32, the learning rate schedule is \emph{adaptive}, the activation is \emph{logistic}, $\alpha$ is $10^{-5}$, the maximum number of iterations is 2000 and the initial learning rate is 0.005.
  \item \textbf{Attention-RNN:} A multi-layer bidirectional RNN network structure with Bahdanau attention mechanism~\citep{Bahdanau14}. After some tuning experiments of parameters, the number of hidden layer is set to 8 and each layer has 128 GRU cells. We set the training parameters as follows: the initial learning rate is 0.005 and the batch size is 128. And we use Adam optimization method to find the best weights in 20 epochs.
 \end{itemize}
  
 \subsubsection{Regression Baselines}
 \textbf{SVR}, \textbf{LR}, \textbf{Ridge}, \textbf{Decision Tree} , \textbf{Random Forest} and \textbf{MLP}: They are same as in classification task, except that their tasks are changed to regression.

 \textbf{Attention-RNN}: Different from the end of attention-RNN in classification, a fully connected layer without softmax activation function is connected at the end of models. We used the MAE loss function during the training. According to our parameter analysis experiment, 8 hidden layers are used, and each layer has 64 GRU cells. Other settings and parameters are consistent with attention-RNN in classification task.

\subsection{Attention Visualization}
\begin{figure}
  \centering
  \includegraphics[width=0.8\columnwidth]{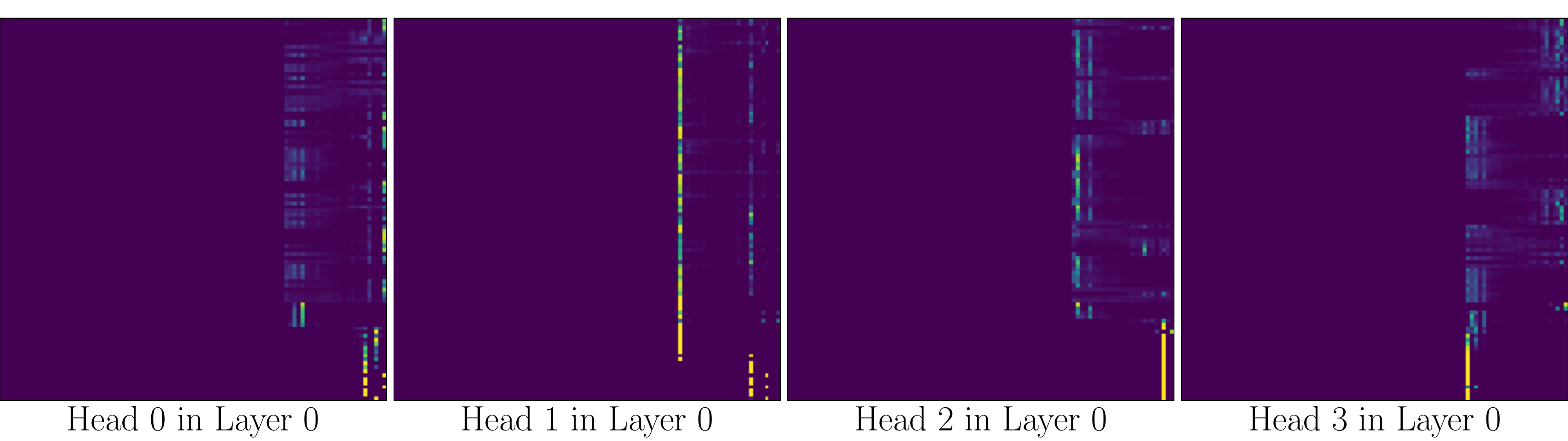}
  \caption{Graphical illustration of attention in the first encode layer.}
  \label{fig:att}
\end{figure}

We plot the weight distribution of the attention which in our Reg-Fellow model. The attention distributions of $head 0$ to $head 3$ in the first Transformer encode layer are shown in Fig.~\ref{fig:att}. 

There are several sub-figures in Fig.~\ref {fig:att}, each of them is a distribution visualization of attention weight for a head. In each sub-figure, the left area representing the period before candidate first published his/her paper is completely blank. The attention in the first Transformer encode layer is mainly concentrated on a few years or even only a year, as shown in Fig.~\ref {fig:att}. However, in deeper encode layers, attention may be distributed more evenly throughout the career of candidate. The relationship between the attention in deep encode layer and the original input is more complicated. Therefore, We only analyze the relationship between the attention in the first encode layer and the original input. For an input sample which represents a candidate $i$, we analysis it as follows: 

\begin{itemize}
  \item First, we extract fragments of the original input, which have greatest attention weights of heads in the first encode layer. The fragments are shown in Eq.~\ref{equ:attention_extract}
  \begin{equation}
    \label{equ:attention_extract}
    \mathbf{A}_{i}=\{\mathbf{a}_{i}^{0},\cdots,\mathbf{a}_{i}^{h},\cdots,\mathbf{a}_{i}^{6}\},
  \end{equation}
  \noindent where each of fragments is a vector (markd $\mathbf{a}_{i}^{h}$, $0$ $\leq$ $h$ $\leq$ $6$) representing the information of the candidate $i$ and $h$ is the index of heads in the first encode layer.
  \item Second, we find the largest elements in each vector $\mathbf{a}_{i}^{h}$, which reveals that candidate $i$ is relatively prominent in this dimension. 
  \item Third, we count the frequency of the dimensions where the largest elements are located.
\end{itemize}

\begin{figure}[ht]
    \centering
  \includegraphics[width=0.7\linewidth]{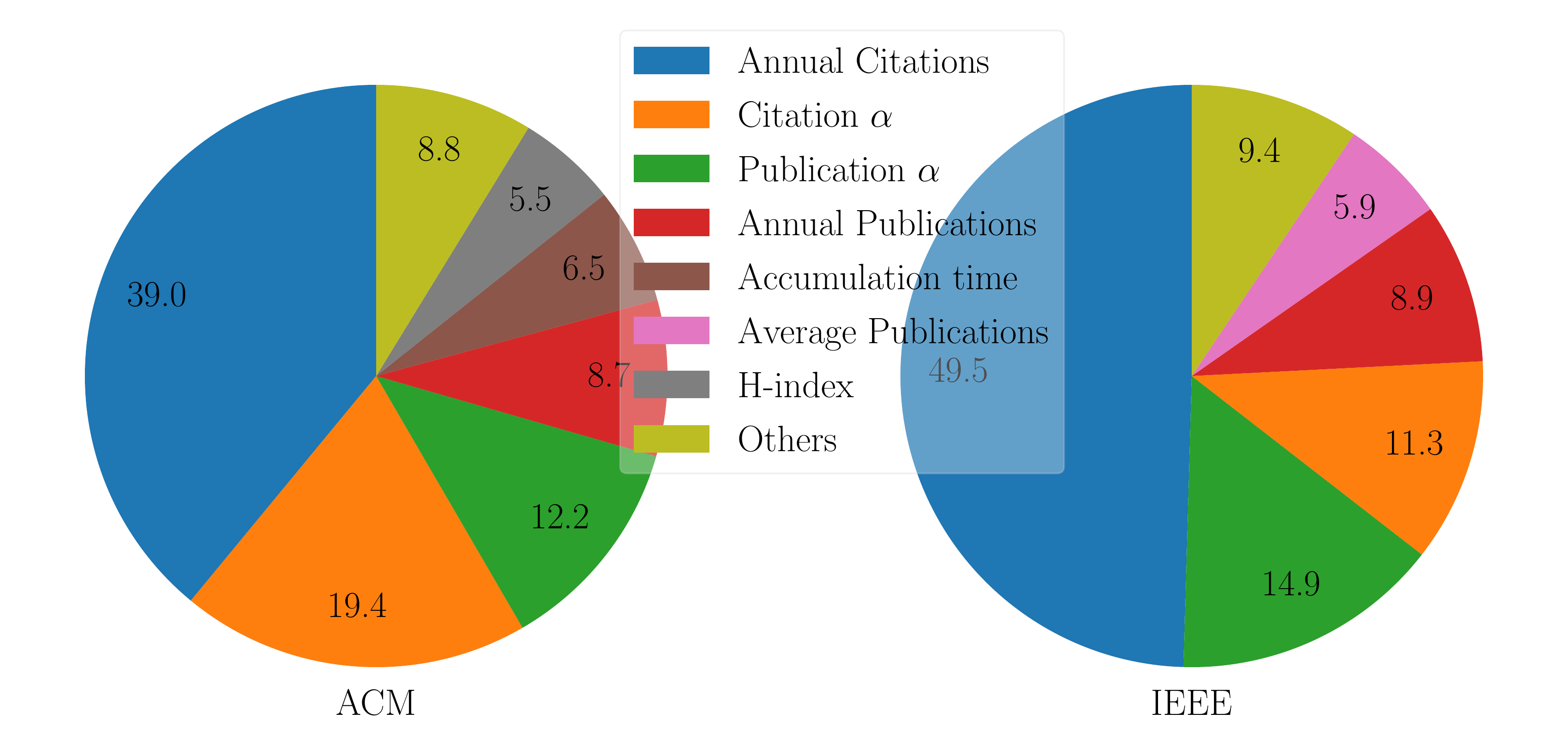}
  \caption{Attention-related factors and their frequency (\%) for ACM/IEEE Fellows}
\label{fig:att_features}
\end{figure}

By analyzing on IEEE regression dataset and ACM regression dataset according to the above steps, we can obtain the factors and their frequencies that are closely related to the attention of the first encode layer, as shown in Fig.~\ref{fig:att_features}. We can find that these two figures are similar: 
\begin{itemize}
  \item The \textbf{Annual citations} have received the most attention in both IEEE and ACM. The higher annual citations means that people are very appreciative of the candidates' research, that is, candidates have greater academic influence, during that period. The Attention of our Reg-Fellow model implies that academic influence in certain periods is crucial for predicting when a candidate will be elected.
  \item \textbf{Publication $\alpha$} and \textbf{Citation $\alpha$} occupy a large proportion in the pie figures, which means scholarly productivity, especially the growth of academic output, also plays an important role.
  \item For accumulation factors, such as \textbf{Total Citations} and \textbf{Accumulation  time}, they are not focused on by the first encode layer in our model. Maybe, they are processed by deeper encode layers.
\end{itemize}

\section{Declarations}
\subsection{Funding}
No funding was received to assist with the preparation of this manuscript.

\subsection{Conflicts of interest}
The authors have no conflicts of interest to declare that are relevant to the content of this article.

\subsection{Availability of data and source code}
The datasets and source code can be downloaded from github repository (\url{https://github.com/nobrowning/Fellow_Analysis}).
We will continue to update them in the future.

\subsection{Ethics approval}
Not applicable.

\subsection{Consent to participate}
Not applicable.

\subsection{Consent for publication}
Not applicable.

%
%

\bibliography{references}

\end{document}